\documentclass[12pt]{article}%
\pdfoutput=1
\usepackage[nosort]{cite}
\usepackage{graphicx}
\usepackage{multicol}
\usepackage{amsfonts}
\usepackage{amssymb}
\usepackage{amsmath}
\usepackage{heck}
\usepackage{afterpage}
\usepackage{setspace}
\usepackage{verbatim}
\usepackage{color}
\usepackage{longtable}
\usepackage{float}
\usepackage{subcaption}
\usepackage{epsfig}
\usepackage{epstopdf}
\usepackage[enableskew, vcentermath]{youngtab}
\usepackage{adjustbox}
\newsavebox{\mysavebox}

\usepackage{tikz}
\usepackage[margin=1in]{geometry}
\usepackage{titletoc}%
\usepackage{hyperref}
\hypersetup{colorlinks,%
citecolor=black,%
filecolor=black,%
linkcolor=black,%
urlcolor=black,%
pdftex}
\providecommand{\U}[1]{\protect\rule{.1in}{.1in}}
\usetikzlibrary{decorations.markings}

\numberwithin{equation}{section}

\newcommand{\ba}{\begin{eqnarray}}
\newcommand{\ea}{\end{eqnarray}}

\newcommand{\be}{\begin{equation}}
\newcommand{\ee}{\end{equation}}

\tikzstyle{startstop} = [rectangle, rounded corners, minimum width=3cm, minimum height=1cm,text centered, draw=black, fill=blue!10]
\tikzstyle{startstop} = [rectangle, rounded corners, minimum width=3cm, minimum height=1cm,text centered, draw=black, fill=blue!10]

\tikzstyle{io} = [trapezium, trapezium left angle=70, trapezium right angle=110, minimum width=3cm, minimum height=1cm, text centered, draw=black, fill=blue!30]

\tikzstyle{process} = [rectangle, minimum width=3cm, minimum height=1cm, text centered, draw=black, fill=orange!30]
\tikzstyle{decision} = [diamond, minimum width=3cm, minimum height=1cm, text centered, draw=black, fill=green!30]

\tikzstyle{arrow} = [thick,->,>=stealth]

\tikzset{->-/.style={decoration={
  markings,
  mark=at position #1 with {\arrow[scale=2.4]{>}}},postaction={decorate}}}

\makeatletter \@addtoreset{equation}{section} \makeatother

\begin{document}

\date{September 2016}

\title{Punctures for Theories of Class $\mathcal{S}_\Gamma$}

\institution{UNC}{\centerline{${}^{1}$Department of Physics, University of North Carolina, Chapel Hill, NC 27599, USA}}

\institution{HARVARD}{\centerline{${}^{2}$Jefferson Physical Laboratory, Harvard University, Cambridge, MA 02138, USA}}

\authors{Jonathan J. Heckman\worksat{\UNC}\footnote{e-mail: {\tt jheckman@email.unc.edu}},
Patrick Jefferson\worksat{\HARVARD}\footnote{e-mail: {\tt jeff@physics.harvard.edu}},\\[4mm]
Tom Rudelius\worksat{\HARVARD}\footnote{e-mail: {\tt rudelius@physics.harvard.edu}},
and Cumrun Vafa\worksat{\HARVARD}\footnote{e-mail: {\tt vafa@physics.harvard.edu}}}

\abstract{With the aim of understanding compactifications of
6D superconformal field theories to four dimensions, we
study punctures for theories of class $\mathcal{S}_{\Gamma}$. The class $\mathcal{S}_{\Gamma}$
theories arise from M5-branes probing $\mathbb{C}^2 / \Gamma$, an ADE singularity.
The resulting 4D theories descend from compactification on Riemann surfaces decorated
with punctures. We show that for class $\mathcal{S}_{\Gamma}$ theories,
a puncture is specified by singular boundary conditions for fields in the 5D quiver gauge theory
obtained from compactification of the 6D theory on a cylinder geometry.
We determine general boundary conditions and study in detail
solutions with first order poles. This yields a generalization of the Nahm pole data present for $1/2$ BPS punctures
for theories of class $\mathcal{S}$. Focusing on specific algebraic structures, we
show how the standard discussion of nilpotent orbits and its connection to representations of $\mathfrak{su}(2)$
generalizes in this broader context.}

\maketitle

\tableofcontents

\enlargethispage{\baselineskip}

\setcounter{tocdepth}{2}

\newpage

\section{Introduction \label{sec:INTRO}}

One of the remarkable developments from recent work on quantum fields and
strings is the close interplay between higher-dimensional
theories and their lower-dimensional compactified descendants. The higher-dimensional
perspective often provides a simple geometric explanation of non-trivial
strongly coupled phenomena in lower dimensions.

From this perspective, it is natural to consider compactifications of 6D
superconformal field theories (SCFTs): six is the largest dimension permitting the existence of an
SCFT \cite{Nahm:1977tg}, and it is tempting to conjecture that all lower-dimensional SCFTs
arise from appropriate compactifications of these \textquotedblleft master
theories.\textquotedblright\ Given the classification of $(2,0)$ and $(1,0)$
6D\ SCFTs via F-theory \cite{Heckman:2013pva,DelZotto:2014hpa,Heckman:2014qba,Heckman:2015bfa},
the time is ripe to ask what new theories can be obtained via compactification to lower dimensions---in particular, four dimensions. This has already been carried out for
the $(2,0)$ theories compactified on Riemann surfaces, leading to 4D
$\mathcal{N}=2$ supersymmetric systems that have been studied extensively
\cite{Gaiotto:2009we} (see also
\cite{Klemm:1996bj,Brandhuber:1997cc,Landsteiner:1997vd,Evans:1997hk,Witten:1997sc,Landsteiner:1997ei,
Kapustin:1998xn,Landsteiner:1998pb,Argyres:2002xc,Alday:2009aq,Chacaltana:2010ks,Nekrasov:2012xe,Xie:2012hs,Chacaltana:2012zy}%
). An important ingredient in this story is the study of Riemann surfaces with
punctures, where the choice of the punctures dramatically impacts the
resulting 4D theory. These punctures are associated with boundary conditions
for operators of the 6D theory extended along a real codimension two subspace (the
noncompact 4D spacetime). The full classification of choices of punctures
for class $\mathcal S$ theories is still incomplete. Nonetheless, a subset called
``regular punctures'' have been classified and are related to homomorphisms $\mathfrak{su}(2) \rightarrow \mathfrak{g}_{ADE}$
for class $\mathcal{S}$ theories of type $\mathfrak{g}_{ADE}$ an ADE Lie algebra \cite{Chacaltana:2012zy}.

In the case of 6D$\ $SCFTs with $\mathcal{N}=(1,0)$ supersymmetry,
compactification on a Riemann surface will generically lead to a
4D$\ \mathcal{N}=1$ supersymmetric theory. Some aspects of these theories have
been studied \cite{Gaiotto:2015usa, DelZotto:2015isa, Franco:2015jna, Hanany:2015pfa,
Ohmori:2015pua, Ohmori:2015pia, DelZotto:2015rca, Aganagic:2015cta, Coman:2015bqq, Morrison:2016nrt}.
Much as in the case of the $(2,0)$ theories, additional data is associated with possible boundary
conditions for fields of the 6D theory, i.e. a choice of punctures on the
compactification manifold.

In this paper we initiate the study of punctures of $\mathcal{N}=(1,0)$ SCFTs.
We focus on the specific case of $N$ M5-branes probing an ADE\ singularity $\mathbb C^2/\Gamma$. In
accord with the nomenclature used for $(2,0)$ theories, we refer to these theories as ``class
$\mathcal{S}_{\Gamma}$,'' where $\Gamma$ is a discrete ADE subgroup of $SU(2)$ indicating the type singularity. For a
preliminary discussion of punctures in the case $\Gamma = \mathbb{Z}_k$, see \cite{Gaiotto:2015usa}.

These 6D theories provide examples of \textquotedblleft conformal
matter\textquotedblright\ \cite{DelZotto:2014hpa}, and form the building blocks for
more elaborate 6D\ SCFTs \cite{Heckman:2015bfa}. Already for this limited class, we find a
much broader class of possible $1/2$ BPS\ punctures than what is obtained for
the $(2,0)$ theories, leading to a rich class of novel 4D theories. We defer
the challenging question of classification to future work.

The basic idea is rather simple: Studying the allowed supersymmetric
punctures is equivalent to specifying supersymmetric boundary conditions for
compactification of these theories on a cylinder, viewed as a semi-infinite
tube sticking out of the Riemann surface. The semi-infinite tube can be viewed as
$S^{1}\times\mathbb{R}_{\geq 0}$. So we first have to study the resulting 5D theory
obtained by compactifying the $(1,0)$ theory on the $S^{1}$ factor, in which
we have some singular behavior for fields in the $\mathbb{R}_{\geq 0}$ factor. For the
class of theories obtained from M5-branes probing $\mathbb{C}^{2}%
/\Gamma$, with $\Gamma \subset SU(2)$ an ADE discrete subgroup,
the resulting 5D system is an affine ADE quiver gauge
theory that admits a Lagrangian description. The gauge algebra is:%
\begin{equation}
\mathfrak{g}_{\text{Quiver}}=\underset{i~\in~\text{Dynkin}}{%
%TCIMACRO{\dprod }%
%BeginExpansion
{\displaystyle\prod}
%EndExpansion
}\mathfrak{u}(Nd_{i}),
\end{equation}
where $N$ is the total number of M5-branes, the product on $i$ runs over the
nodes of the corresponding affine ADE Dynkin diagram, and $d_{i}$ is the Dynkin
index of a node in the graph. The links between these gauge groups are 5D
$\mathcal{N}=1$ hypermultiplets in bifundamental representations. See Table \ref{tab:quiver}
for a depiction of the associated quiver gauge theories for each of the ADE subgroups. We use this Lagrangian description to
determine the allowed supersymmetric boundary conditions for fields of the
quiver theory with poles at the origin of $\mathbb{R}_{\geq 0}$. In this work we primarily focus on
the case of fields with simple poles: regular punctures.

In the special case where $\Gamma$ is trivial, we recover the
punctures of a $(2,0)$ theory. However, since we only demand that four
real supercharges are preserved, this already leads us to $1/4$ BPS\ punctures
of the $(2,0)$ theory. These boundary conditions are characterized by the
equations:%
\begin{equation}
\lbrack\Sigma,Q]=Q\text{, \ \ }[\Sigma,\widetilde{Q}]=\widetilde{Q}\text{,
\ \ }[Q,\widetilde{Q}]=0\text{, \ \ }[Q,Q^{\dag}]+[\widetilde{Q}%
,\widetilde{Q}^{\dag}]=\Sigma, \label{GenNahm}%
\end{equation}
where $\Sigma$, $Q$ and $\widetilde{Q}$ are $N \times N$ matrices with complex entries, with
$\Sigma$ Hermitian. The special case of $1/2$ BPS\ punctures for a $(2,0)$ theory is recovered by
setting $\widetilde{Q}=0$, for which the above system reduces to the commutation relations specifying a representation of $\mathfrak{su}(2)$.
These conditions may equivalently be viewed as determining a
nilpotent element $Q$ in the simply-laced algebra $\mathfrak{g}$ of the $(2,0)$ theory in
question. Equation (\ref{GenNahm}), with $\widetilde Q \ne 0$, is the natural generalization of this. As
we show, these equations specify a pair of commuting nilpotent matrices $Q$ and
$\widetilde{Q}$ subject to additional constraints. For earlier work on $1/4$ BPS punctures for theories of class $\mathcal{S}$,
see \cite{Xie:2013gma}.

For class $\mathcal{S}_{\Gamma}$ theories, a $1/2$ BPS\ puncture preserves
four real supercharges. The boundary conditions we find are most conveniently
stated in terms of an algebra of $N\left\vert \Gamma\right\vert
\times N\left\vert \Gamma\right\vert $ matrices with entries in $%
%TCIMACRO{\U{2102} }%
%BeginExpansion
\mathbb{C}
%EndExpansion
$, where $\left\vert \Gamma\right\vert $ is the order of the discrete ADE
subgroup $\Gamma\subset SU(2)$. Given $\Sigma$ Hermitian and $Q$ and
$\widetilde{Q}$ matrices with general complex entries, the set of regular punctures
$\mathcal{P}$ obeys the conditions of equation (\ref{GenNahm}). To get a
solution for the quiver gauge theory, we project to
the quiver basis of fields as dictated by the Douglas-Moore orbifold
construction \cite{Douglas:1996sw}, retaining only $\Gamma$-equivariant solutions
$\mathcal{P}_{\Gamma}$.

Now, in the case of the A-type $(2,0)$ theories, there is a beautiful
characterization of punctures in terms of nilpotent orbits of
$\mathfrak{u}(N)$, or equivalently Young diagrams with $N$ boxes. By a theorem of
Jacobson-Morozov, these are in one-to-one correspondence with homomorphisms
$\mathfrak{su}(2)\rightarrow\mathfrak{u}(N)$. Similar considerations hold for
the other $(2,0)$ theories, where $\mathfrak{u}(N)$ is instead replaced by a
different choice of ADE Lie algebra $\mathfrak{g}_{ADE}$.

It is natural to ask how this characterization generalizes to $(1,0)$
theories. Perhaps the closest analogue of the standard Nahm pole equations
comes from taking $\widetilde{Q}=0$, but with $\Gamma$ non-trivial. Here, we
obtain a full classification of possible punctures in terms of Young
diagrams decorated by appropriate roots of unity. In the case where $Q$ and
$\widetilde{Q}$ define a pair of commuting $\mathfrak{su}(2)$s, we again obtain a full
classification of solutions. An interesting feature of these solutions is that
only in the A- and D-type quivers do we obtain non-trivial solutions.
More broadly, we also find a partial
characterization of solutions with a product of $\mathfrak{su}(2)$s:
\begin{equation}
\underset{l}{\underbrace{\mathfrak{su}(2)\times...\times\mathfrak{su}(2)}%
}\rightarrow\mathcal{P}\rightarrow\mathcal{P}_{\Gamma}%
\end{equation}
for $l$ some number of $\mathfrak{su}(2)$ factors. These are combinatorially
represented in terms of self-avoiding directed paths through the corresponding ADE\ quiver.

The rest of the paper is organized as follows. In section \ref{sec:NAHM}, we
review some aspects of Nahm pole data for the $(2,0)$ theories, and then
present a generalization to the case of M5-branes probing an
ADE\ singularity. Section \ref{sec:PAIRS} contains remarks about the fact that
the generalized Nahm pole equations involve a pair of commuting nilpotent matrices.
In section \ref{sec:SU2}, we turn to the closest
analogue of the (2,0) solutions, namely those for which $\widetilde{Q}=0$. In section \ref{sec:TWOSU2},
we provide a mild generalization where $Q$ and $\widetilde{Q}$ generate
an $\mathfrak{su}(2)_{Q} \times \mathfrak{su}(2)_{\widetilde{Q}}$ algebra.
We then turn in section \ref{sec:GRAPH} to solutions based on directed
self-avoiding paths. We present our conclusions and directions for future work
in section \ref{sec:CONC}. Additional low rank examples supplementing the discussion can be found in Appendix \ref{app:FURTHER}.

\begin{table}
	\begin{center}
	$
	\begin{array}{|c|c|}\hline \text{ADE type}& \text{Quiver}  \\   \hline
		\begin{tikzpicture} \node at (0,1.4) {$\widehat{A}_k$};
		\node at (0,0) {};
		\end{tikzpicture} &\begin{tikzpicture}[scale=.9]
			\node[draw,circle,thick,dotted] (top) at (0,2.5) {$N$};
			\node[draw,circle,thick,dotted] (ll) at (-5,0) {$N$};
			\node[draw,circle,thick,dotted] (l) at (-2.5,0) {$N$};
			\node[] (c) at (0,0) {$\dots \dots$};
			\node[draw,circle,thick,dotted] (rr) at (5,0) {$N$};
			\node[draw,circle,thick,dotted] (r) at (2.5,0) {$N$};
			\draw[->-=.5] (top) to [bend right = 6] (ll);
			\draw[->-=.5] (ll) to [bend right = 6] (top);
			\draw[->-=.5] (ll) to [bend right = 10]  (l);
			\draw[->-=.5] (l) to [bend right = 10]  (ll);
			\draw[->-=.5] (l) to  [bend right = 10]  (c);
			\draw[->-=.5] (c) to  [bend right = 10]  (l);
			\draw[->-=.5] (c) to  [bend right = 10]  (r);
			\draw[->-=.5] (r) to  [bend right = 10]  (c);
			\draw[->-=.5] (r) to [bend right = 10] (rr);
			\draw[->-=.5] (rr) to [bend right = 6]   (top);
			\draw[->-=.5] (rr) to [bend right = 10] (r);
			\draw[->-=.5] (top) to [bend right = 6]   (rr);
			%\node at (-8,0) {type $\widehat{A}_k$};
		\end{tikzpicture}	 \\\hline
	\begin{tikzpicture}
		\node at (0,1.8) {$	\widehat{D}_k$};
		\node at (0,0) {};
	\end{tikzpicture}
		&\begin{tikzpicture}[scale=.9,yscale=.9]
			\node[draw,circle,thick,dotted] (a) at (-5,2) {$N$};
			\node[draw,circle,thick,dotted] (b) at (-5,-2) {$N$};
			\node[draw,circle,thick,dotted] (cc) at (5,2) {$N$};
			\node[draw,circle,thick,dotted] (d) at (5,-2) {$N$};
			\node[draw,circle,thick,dotted,scale=.85] (ll) at (-3.5,0) {$2N$};
			\node[draw,circle,thick,dotted,scale=.85] (l) at (-1.5,0) {$2N$};
			\node[] (c) at (0,0) {$ \dots$};
			\node[draw,circle,thick,dotted,scale=.85] (rr) at (3.5,0) {$2N$};
			\node[draw,circle,thick,dotted,scale=.85] (r) at (1.5,0) {$2N$};
			\draw[->-=.5] (ll) to [bend right = 10]  (l);
			\draw[->-=.5] (l) to [bend right = 10]  (ll);
			\draw[->-=.5] (l) to  [bend right = 10]  (c);
			\draw[->-=.5] (c) to  [bend right = 10]  (l);
			\draw[->-=.5] (c) to  [bend right = 10]  (r);
			\draw[->-=.5] (r) to  [bend right = 10]  (c);
			\draw[->-=.5] (r) to [bend right = 10] (rr);
			\draw[->-=.5] (rr) to [bend right = 10] (r);
			\draw[->-=.5] (a) to [bend right = 10] (ll);
			\draw[->-=.5] (ll) to [bend right = 10] (a);
			\draw[->-=.5] (b) to [bend right = 10] (ll);
			\draw[->-=.5] (ll) to [bend right = 10] (b);
			\draw[->-=.5] (cc) to [bend left = 10] (rr);
			\draw[->-=.5] (rr) to [bend left = 10] (cc);
			\draw[->-=.5] (d) to [bend left = 10] (rr);
			\draw[->-=.5] (rr) to [bend left = 10] (d);
			%\node at (-7,0) {type $\widehat{D}_k$};
		\end{tikzpicture}	\\\hline
		\begin{tikzpicture}
		\node at (0,2) {$\widehat{E}_6$};
		\node at (0,0) {};
		\end{tikzpicture}
		&\begin{tikzpicture}[scale=1.2]
			\node[draw,circle,thick,dotted] (ll) at (-3,0) {$N$};
			\node[draw,circle,thick,dotted,scale=.85] (l) at (-1.5,0) {$2N$};
			\node[draw,circle,thick,dotted,scale=.85] (c) at (0,0) {$3N$};
			\node[draw,circle,thick,dotted,scale=.85] (r) at (1.5,0) {$2N$};
			\node[draw,circle,thick,dotted] (rr) at (3,0) {$N$};
			\node[draw,circle,thick,dotted,scale=.85] (u) at (0,1.5) {$2N$};
				\node[draw,circle,thick,dotted] (uu) at (0,3) {$N$};
				\draw[->-=.5] (ll) to [bend right = 10] (l);
				\draw[->-=.5] (l) to [bend right = 10] (ll);
				\draw[->-=.5] (l) to  [bend right = 10] (c);
				\draw[->-=.5] (c) to  [bend right = 10] (r);
				\draw[->-=.5] (r) to [bend right = 10] (rr);
				\draw[->-=.5] (c) to [bend right = 10] (u);
				\draw[->-=.5] (u) to [bend right = 10] (uu);
				\draw[->-=.5] (c) to  [bend right = 10] (l);
				\draw[->-=.5] (r) to  [bend right = 10] (c);
				\draw[->-=.5] (rr) to [bend right = 10] (r);
				\draw[->-=.5] (u) to [bend right = 10] (c);
				\draw[->-=.5] (uu) to [bend right = 10] (u);
				%\node at (-4.5,0) {type $\widehat{E}_6$};
		\end{tikzpicture} \\\hline
		\begin{tikzpicture}
			\node at (0,1.2) {$\widehat{E}_7$};
			\node at (0,0) {};
		\end{tikzpicture}
		&\begin{tikzpicture}[scale=1.2]
			\node[draw,circle,thick,dotted] (lll) at (-4.5,0) {$N$};
			\node[draw,circle,thick,dotted,scale=.85] (ll) at (-3,0) {$2N$};
			\node[draw,circle,thick,dotted,scale=.85] (l) at (-1.5,0) {$3N$};
			\node[draw,circle,thick,dotted,scale=.85] (c) at (0,0) {$4N$};
			\node[draw,circle,thick,dotted,scale=.85] (r) at (1.5,0) {$3N$};
			\node[draw,circle,thick,dotted,scale=.85] (rr) at (3,0) {$2N$};
			\node[draw,circle,thick,dotted] (rrr) at (4.5,0) {$N$};
			\node[draw,circle,thick,dotted,scale=.85] (u) at (0,1.5) {$2N$};
				\draw[->-=.5] (ll) to [bend right = 10] (l);
				\draw[->-=.5] (l) to [bend right = 10] (ll);
				\draw[->-=.5] (l) to  [bend right = 10] (c);
				\draw[->-=.5] (c) to  [bend right = 10] (r);
				\draw[->-=.5] (r) to [bend right = 10] (rr);
				\draw[->-=.5] (c) to [bend right = 10] (u);
				\draw[->-=.5] (c) to  [bend right = 10] (l);
				\draw[->-=.5] (r) to  [bend right = 10] (c);
				\draw[->-=.5] (rr) to [bend right = 10] (r);
				\draw[->-=.5] (u) to [bend right = 10] (c);
				\draw[->-=.5] (lll) to [bend right = 10] (ll);
				\draw[->-=.5] (ll) to [bend right = 10] (lll);
				\draw[->-=.5] (rrr) to [bend right = 10] (rr);
				\draw[->-=.5] (rr) to [bend right = 10] (rrr);
				%	\node at (-6,0) {type $\widehat{E}_7$};
		\end{tikzpicture} \\\hline
		\begin{tikzpicture}
			\node at (0,1.2) {$\widehat{E}_8$};
			\node at (0,0) {};
		\end{tikzpicture}
		&\begin{tikzpicture}[scale=1.2]
			\node[draw,circle,thick,dotted,scale=.85] (lll) at (-4.5,0) {$2N$};
			\node[draw,circle,thick,dotted,scale=.85] (ll) at (-3,0) {$4N$};
			\node[draw,circle,thick,dotted,scale=.85] (l) at (-1.5,0) {$6N$};
			\node[draw,circle,thick,dotted,scale=.85] (c) at (0,0) {$5N$};
			\node[draw,circle,thick,dotted,scale=.85] (r) at (1.5,0) {$4N$};
			\node[draw,circle,thick,dotted,scale=.85] (rr) at (3,0) {$3N$};
			\node[draw,circle,thick,dotted,scale=.85] (rrr) at (4.5,0) {$2N$};
			\node[draw,circle,thick,dotted,scale=.85] (u) at (-1.5,1.5) {$3N$};
			\node[draw,circle,thick,dotted] (rrrr) at (6,0) {$N$};
				\draw[->-=.5] (ll) to [bend right = 10] (l);
				\draw[->-=.5] (l) to [bend right = 10] (ll);
				\draw[->-=.5] (l) to  [bend right = 10] (c);
				\draw[->-=.5] (c) to  [bend right = 10] (r);
				\draw[->-=.5] (r) to [bend right = 10] (rr);
				\draw[->-=.5] (l) to [bend right = 10] (u);
				\draw[->-=.5] (c) to  [bend right = 10] (l);
				\draw[->-=.5] (r) to  [bend right = 10] (c);
				\draw[->-=.5] (rr) to [bend right = 10] (r);
				\draw[->-=.5] (u) to [bend right = 10] (l);
				\draw[->-=.5] (lll) to [bend right = 10] (ll);
				\draw[->-=.5] (ll) to [bend right = 10] (lll);
				\draw[->-=.5] (rrr) to [bend right = 10] (rr);
				\draw[->-=.5] (rr) to [bend right = 10] (rrr);
				\draw[->-=.5] (rrrr) to [bend right = 10] (rrr);
				\draw[->-=.5] (rrr) to [bend right = 10] (rrrr);
					%\node at (-4.3,1) {type $\widehat{E}_8$};
		\end{tikzpicture}
		\\\hline
	\end{array}
	$
	\end{center}
	\caption{Punctures for M5-branes probing an ADE singularity are specified in terms of Nahm pole data in an associated 5D gauge theory obtained by reduction on a circle. This is a 5D
a quiver gauge theory with nodes and links fields specified by the corresponding affine ADE Dynkin diagram. Each quiver node has gauge group $U(d_i N)$ where $d_i$ is the Dynkin index of the node and $N$ is the total number of M5-branes probing the singularity.}
	\label{tab:quiver}
\end{table}

\section{Punctures and M5-branes \label{sec:NAHM}}

In this section we introduce the primary class of theories for which we will
study punctures. These are given by M5-branes probing an ADE\ singularity,
i.e., we consider spacetime-filling branes in which the transverse geometry is
$\mathbb{R}_{\bot}\times\mathbb{C}^{2}/\Gamma$, with $\Gamma$ discrete
subgroup of $SU(2)$. To study the structure of punctures in this theory, we
then partially compactify on a cylinder $\mathbb{C}^{\ast}$ so that the full
geometry is of the form $\mathbb{R}^{3,1}\times\mathbb{C}^{\ast}%
\times\mathbb{R}_{\bot}\times\mathbb{C}^{2}/\Gamma$. Our goal will be to
understand boundary conditions associated with the cylinder geometry that
preserve four real supercharges. That is, we will be left with a 4D system
with $\mathcal{N} = 1$ supersymmetry. We present a general analysis of singular field
profiles, but shall primarily focus on the case of fields with first order poles,
i.e., the case of regular punctures. For early work on $1/2$ BPS boundary conditions
and its connection to the Nahm pole equations see e.g.
\cite{Diaconescu:1996rk, Tsimpis:1998zh, Kapustin:1998pb, DeWolfe:2001pq, Erdmenger:2002ex,
Constable:2002xt, Gaiotto:2008sa}.

The primary strategy we adopt to study this question is to recognize that
topologically the cylinder $\mathbb{C}^{\ast}$ is simply given by $S^{1}%
\times\mathbb{R}$. Since the circle reduction of M5-brane theories leads to a
5D Lagrangian field theory (with a UV cutoff), we can equally well study
boundary conditions in the 5D theory on the factor $\mathbb{R}$.

The rest of this section is organized as follows. First, we review the
standard analysis of punctures in the special case where $\Gamma$ is trivial,
which brings the discussion into contact with compactifications of the A-type 6D
$(2,0)$ SCFTs. We then turn to the analogous question for non-trivial $\Gamma$.
We determine supersymmetric boundary conditions preserving four
supercharges in the presence of a real codimension two defect. Using these conditions, we
then derive a system of algebraic equations that must be satisfied by a puncture.

\subsection{$1/4$ BPS\ Punctures for Class $\mathcal{S}$ Theories}

Let us now turn to an analysis of punctures in the $(2,0)$ theories which
preserve four real supercharges, i.e., the case of $1/4$ BPS\ punctures. Although our
main focus will be the A-type $(2,0)$ theories realized geometrically by stack of $N$ M5-branes
in flat space, the results described in this subsection readily generalize
to the other ADE $(2,0)$ theories.

Reducing the 6D theory on a circle leads to a 5D $\mathcal{N}=2$ supersymmetric gauge
theory with gauge group $U(N)$. There are various ways to determine boundary
conditions on a cylinder which preserve some fraction of the bulk
supersymmetry. One method is to consider the bosonic equations of motion obtained by
varying the the 5D $\mathcal{N}=2$ gauginos, and to then impose singular
behavior for some of the fields. An equivalent method is to treat the
higher-dimensional theory in terms of a collection of 4D fields in which we
only impose the standard supersymmetric\ equations of motion for the 4D
theory. This will lead us to boundary conditions which preserve four real supercharges.

Indeed, since we are interested in possible boundary conditions which preserve
a 4D Lorentz invariant vacuum with $\mathcal{N}=1$ supersymmetry, much as in
reference \cite{ArkaniHamed:2001tb}, it is
helpful to assemble the mode content of this 5D theory in terms of a
collection of $\mathcal{N}=1$ multiplets parameterized by points of the factor
$\mathbb{R}$ of the cylinder $S^{1}\times\mathbb{R}$.
With this in mind, we
have a collection of 4D vector multiplets, and three adjoint-valued chiral
multiplets, all of which are labelled by internal points of $\mathbb{R}$. One
of these chiral multiplets transforms as a vector on $S^{1}\times\mathbb{R}$,
so we denote it by $Z(t)$, while the other two arrange as $Q(t)\oplus \widetilde{Q}(t)$, a
collection of 4D $\mathcal{N}=2$ hypermultiplets, which transforms as a scalar on
$\mathbb{R}$. It is helpful to further decompose $Z(t)$ locally as the
complexified connection:%
\begin{equation}
Z(t)=\partial_{t}+ \frac{1}{\sqrt{2}}\left( \Sigma(t) + i A_{t} \right)  , \label{compconn}%
\end{equation}
where $t$ is the coordinate along $\mathbb{R}$ (with connection $A_{t}$) and
$\Sigma(t)$ is the adjoint-valued real scalar in a 5D $\mathcal{N}=1$ vector
multiplet. Note that by a suitable choice of gauge, we can locally set $A_{t}=0$.
One should keep in mind that on a topologically non-trivial Riemann surface, this
is not possible to do globally. Geometrically, $\Sigma(t),Q(t)$ and
$\widetilde{Q}(t)$ rotate as a vector of $SO(5)$, the R-symmetry group of the
$(2,0)$ theory.

The BPS\ equations of motion obtained from the condition that we have a
Lorentz invariant 4D $\mathcal{N}=1$ vacuum are:
\begin{align}
\text{F-terms}  &  \text{: }[Z(t),Q(t)]=[Z(t),\widetilde{Q}%
(t)]=[Q(t),\widetilde{Q}(t)]=0\\
\text{D-term}  &  \text{: }[Z(t),Z^{\dag}(t)]+[Q(t),Q^{\dag}%
(t)]+[\widetilde{Q}(t),\widetilde{Q}^{\dag}(t)]=0,
\end{align}
modulo $\mathfrak{u}(N)$ gauge transformations. Note that the commutator with the $Z$'s is just
an internal field strength:
\begin{equation}
[Z(t) , Z^{\dag}(t)] = \partial_{t} \Sigma(t).
\end{equation}

Now, we are interested in possibly non-trivial boundary conditions for our
fields along the factor of $\mathbb{R}$. By a change of coordinates, we can
take this singularity to lie at $t=0$, and we consider the case where the
fields have simple poles:%
\begin{equation}
Q(t)=\frac{Q}{t},\text{ \ \ }\widetilde{Q}(t)=\frac{\widetilde{Q}}{t},\text{
\ \ }\Sigma(t)=\frac{\Sigma}{t}.
\end{equation}
Observe that since $Z=\partial_{t} + \Sigma(t) / \sqrt{2}$ (in the gauge $A_{t}=0$), both
components can act by non-trivial commutator on the other fields. Our F-term
and D-term equations of motion thus reduce to:%
\begin{align}
\text{F-term}  &  \text{: }[\Sigma,Q]=Q\\
\text{F-term}  &  \text{: }[\Sigma,\widetilde{Q}]=\widetilde{Q}\\
\text{F-term}  &  \text{: }[Q,\widetilde{Q}]=0\\
\text{D-term}  &  \text{:\ }[Q,Q^{\dag}]+[\widetilde{Q},\widetilde{Q}^{\dag
}]=\Sigma\text{.}%
\end{align}
This is the same generalization of the Nahm pole equations found
in reference \cite{Xie:2013gma} (see also \cite{Hashimoto:2014vpa}).

Now, when $Q,\widetilde{Q}$ and $\Sigma$ are generic, we get a $1/4$
BPS\ puncture retaining four real supercharges. If, however, some
linear combination of $Q$ and $\widetilde{Q}^{\dag}$ vanishes, we retain an $SU(2)$
R-symmetry subgroup of $SO(5)$, preserving a 4D $\mathcal{N}=2$ subalgebra.

The familiar conditions for Nahm poles arise from setting $\widetilde{Q}$ to zero:
\begin{equation}
\lbrack\Sigma,Q]=Q\text{ \ \ and \ \ }\Sigma=[Q,Q^{\dag}]\text{.}%
\end{equation}
Since these algebraic relations define an $\mathfrak{su}(2)$ subalgebra of
$\mathfrak{u}(N)$, we see that such Nahm pole data is captured by a choice of
nilpotent orbit of $\mathfrak{u}(N)$, i.e., a choice of partition / Young diagram.

One can also generalize our discussion to the case of higher order
singularities. Introducing an expansion of the form:%
\begin{equation}
Q(t)=\underset{n>0}{%
%TCIMACRO{\dsum }%
%BeginExpansion
{\displaystyle\sum}
%EndExpansion
}\frac{Q_{n}}{t^{n}},\text{ \ \ }\widetilde{Q}(t)=\underset{n>0}{%
%TCIMACRO{\dsum }%
%BeginExpansion
{\displaystyle\sum}
%EndExpansion
}\frac{\widetilde{Q}_{n}}{t^{n}},\text{ \ \ }\Sigma(t)=\underset{n>0}{%
%TCIMACRO{\dsum }%
%BeginExpansion
{\displaystyle\sum}
%EndExpansion
}\frac{\Sigma_{n}}{t^{n}}.
\end{equation}
In this case, we collect all terms of the same order and demand that they
satisfy the F- and D-term equations of motion:%
\begin{align}
\text{F-term}  &  \text{: }\underset{k+l=m}{\sum}[\Sigma_{k},Q_{l}%
]=(m-1)Q_{m-1}\label{highfirst}\\
\text{F-term}  &  \text{: }\underset{k+l=m}{\sum}[\Sigma_{k},\widetilde{Q}%
_{l}]=(m-1)\widetilde{Q}_{m-1}\\
\text{F-term}  &  \text{: }\underset{k+l=m}{\sum}[Q_{k},\widetilde{Q}_{l}]=0\\
\text{D-term}  &  \text{: }\underset{k+l=m}{\sum}[Q_{k},Q_{l}^{\dag
}]+[\widetilde{Q}_{k},\widetilde{Q}_{l}^{\dag}]=(m-1)\Sigma_{m-1},
\label{highlast}%
\end{align}
for all $k,l,m>0$. Again, we can specialize to $1/2$ BPS\ punctures by setting
$\widetilde{Q}_{k}=0$ for all $k$.

\subsection{Punctures for Class $\mathcal{S}_\Gamma$ Theories}

Having discussed some basic features of the Nahm pole equations for M5-branes
in flat space, we now turn to the analogous set of equations when these branes
probe an ADE\ singularity, namely, the class $\mathcal{S}_{\Gamma}$ theories. This will
realize a 6D SCFT with $\mathcal{N} = (1,0)$ supersymmetry.

In preparation for our analysis of solutions, we shall actually find it
convenient to give two different presentations of the same system of
equations. We refer to these as the \textquotedblleft covering
space\textquotedblright\ basis and the \textquotedblleft quiver
basis,\textquotedblright\ for reasons which will soon be apparent.

Reduction of the M5-brane theory on a circle yields, at low energies, a stack
of D4-branes in type IIA string theory. These D4-branes probe the
ADE\ singularity, leading to a quiver gauge theory that can be derived from
the Douglas-Moore orbifold construction \cite{Douglas:1996sw} (see also \cite{Johnson:1996py, Lawrence:1998ja}).
As we shall be making heavy use of it later, let us briefly review the elements of this
construction. The mode content for the $N$ D4-branes consists of five real
scalars, which are again given by $\Sigma$, $Q$ and $\widetilde{Q}$, but which
now transform in the adjoint representation of $U(N\left\vert \Gamma
\right\vert )$. Viewed as components of a vector on the geometry
$\mathbb{R}\times\mathbb{C}^{2}$, $\Sigma$ is neutral under the $SU(2)$ group
action on $\mathbb{C}^{2}$ while $Q$ and $\widetilde{Q}$ transform as a
doublet, which we write as a two-component vector:%
\begin{equation}
\overrightarrow{Q}=\left[
\begin{array}
[c]{c}%
Q\\
\widetilde{Q}%
\end{array}
\right]  .
\end{equation}

To track the group action of $\Gamma$ on these fields, it is helpful to
decompose the vector space $\mathbb{C}^{N\left\vert \Gamma\right\vert }$ as:%
\begin{equation}
\mathbb{C}^{N\left\vert \Gamma\right\vert }=\underset{i}{\bigoplus}%
\mathbb{C}^{Nd_{i}}\otimes V_{i}, \label{CNG}%
\end{equation}
where here, $i$ runs over the irreducible representations of the discrete
group $\Gamma$, and $d_{i}$ is the dimension of $V_{i}$, which, by the McKay
correspondence, is also the Dynkin index of the corresponding node in the
ADE\ graph. For $\gamma\in\Gamma$, denote by $\rho_{\text{doub}}(\gamma)$ the
$2\times2$ matrix representative, and $\rho_{\text{reg}}(\gamma)$ the regular
representation, i.e. the one which acts on%
\begin{equation}
V_{\text{reg}}=\underset{i}{\bigoplus}\mathbb{C}^{d_i} \otimes V_{i}.
\end{equation}
This canonically extends to a group action on $\mathbb{C}^{N\left\vert
\Gamma\right\vert }$ as in line (\ref{CNG}), so by abuse of notation we also
denote this by $\rho_{\text{reg}}(\gamma)$. The orbifold projection then
amounts to the conditions:%
\begin{equation} \label{orbproj}
\Sigma =\rho_{\text{reg}}(\gamma)\Sigma\rho_{\text{reg}}(\gamma^{-1})
\text{\,\,\, and \,\,\,}
\rho_{\text{doub}}(\gamma)\left[
\begin{array}
[c]{c}%
Q\\
\widetilde{Q}%
\end{array}
\right]   =\left[
\begin{array}
[c]{c}%
\rho_{\text{reg}}(\gamma)Q\rho_{\text{reg}}(\gamma^{-1})\\
\rho_{\text{reg}}(\gamma)\widetilde{Q}\rho_{\text{reg}}(\gamma^{-1})
\end{array}
\right]  .
\end{equation}
To avoid overloading the notation, in what follows we shall often drop the
overall designation of the representation $\rho$ since it will be clear from
the context.

So in other words, punctures of the orbifold theory are obtained by first
imposing the conditions:%
\begin{align}
\text{F-term}  &  \text{: }[\Sigma,Q]=Q\\
\text{F-term}  &  \text{: }[\Sigma,\widetilde{Q}]=\widetilde{Q}\\
\text{F-term}  &  \text{: }[Q,\widetilde{Q}]=0\\
\text{D-term}  &  \text{: }[Q,Q^{\dag}]+[\widetilde{Q},\widetilde{Q}^{\dag
}]=\Sigma\text{,}%
\end{align}
and then imposing the orbifold projection condition of line (\ref{orbproj}).
We refer to this as the \textquotedblleft covering space
basis,\textquotedblright\ since all solutions are embedded in large
$N\left\vert \Gamma\right\vert \times N\left\vert \Gamma\right\vert $ matrices.

Alternatively, we can work in terms of the \textquotedblleft quiver
basis,\textquotedblright\ by directly considering punctures in the
5D\ gauge theory defined by the Douglas-Moore construction. In either case, we
have a product of gauge algebras%
\begin{equation}
\mathfrak{g}_{\text{Quiver}}=\underset{i~\in~\text{Dynkin}}{%
%TCIMACRO{\dprod }%
%BeginExpansion
{\displaystyle\prod}
%EndExpansion
}\mathfrak{u}(Nd_{i}),
\end{equation}
each with gauge coupling \cite{Lawrence:1998ja}:
\begin{equation}\label{tunecoup}
\frac{1}{g_{(i)}^{2}}=\frac{d_{i}}{\left\vert \Gamma\right\vert }\frac
{1}{g_{(5D)}^{2}},
\end{equation}
where as in the usual discussion of compactifications of the $(2,0)$ theory,
the 5D gauge coupling is related to the compactification radius $L$ as
$g_{(5D)}^{2}\sim L$. In the context of the 5D field theory, we are of course free to move away from the
special values dictated by equation (\ref{tunecoup}). In the 4D field theory, the
complexification of these parameters then become marginal parameters. Indeed,
we shall often take convenient values of these couplings when we turn to quiver basis solutions.

Now, for each quiver node $i$, we have an adjoint-valued
field $\Sigma_{i}$, and between pairs $(i,j)$ connected in the Dynkin diagram,
we have a hypermultiplet $H_{(i,j)}\oplus H_{(i,j)}^{c}$, with $H_{(i,j)}$ in
the representation $(Nd_{i},\overline{Nd_{j}})$ and $H_{(i,j)}^{c}$ in the
conjugate representation. Let us stress that in passing from the covering
space basis to the quiver basis, components of $Q$ can contribute to both $H$
and $H^{c}$, and similarly for $\widetilde{Q}$.

To determine the F- and D-term constraints for this system, it is helpful to
recall the superpotential for a 4D, $\mathcal{N} = 2$ supersymmetric gauge theory in a
basis of fields where the scalar of the vector multiplets are canonically
normalized:%
\begin{equation}
W=\underset{i}{\sum}\sqrt{2}g_{(i)}H_{(i,j)}^{c}Z_{(i)}H_{(i,j)}
\label{superpot}%
\end{equation}
where we have introduced the complexified connection $Z_{(i)}$ for each gauge
group factor, as per our discussion near line (\ref{compconn}). To keep the
presentation of F-terms as close to the $(2,0)$ case as possible, it is
convenient to rescale each $Z_{(i)}\rightarrow Z_{(i)}\sqrt{d_{i}}$. In this
rescaled basis of fields, the F- and D-term equations of motion are given by:%
\begin{align}
\text{F-term}  &  \text{: }Z_{(i)}(t)H_{(i,j)}(t)-H_{(i,j)}(t)Z_{(j)}(t)=0\\
\text{F-term}  &  \text{: }H_{(i,j)}^{c}(t)Z_{(i)}(t)-Z_{(j)}(t)H_{(i,j)}%
^{c}(t)=0\\
\text{F-term}  &  \text{: }\underset{j}{\sum}H_{(i,j)}(t)\cdot H_{(i,j)}%
^{c}(t)=0,\text{ \ \ for all }i\\
\text{D-term}  &  \text{: }d_{i}[Z_{(i)}(t),Z_{(i)}^{\dag}%
(t)]+\underset{j}{\sum}\left(  H_{(i,j)}(t)\cdot H_{(i,j)}^{\dag}%
(t)-H_{(i,j)}^{c\dag}(t)\cdot H_{(i,j)}^{c\dag}(t)\right)  =0,\text{ \ \ for
all }i,
\end{align}
where in the above, each pairing $A\cdot B$ is implicitly associated with
the outer product of the fundamental and anti-fundamental representation of
the gauge group $U(Nd_{i})$. We can of course also work out the structure of
the Nahm pole equations in this basis, obtaining the analogous conditions for regular punctures:
\begin{align}
\text{F-term}  &  \text{: }\Sigma_{(i)}H_{(i,j)}-H_{(i,j)}\Sigma
_{(j)}=H_{(i,j)}\\
\text{F-term}  &  \text{: }H_{(i,j)}^{c}\Sigma_{(i)}-\Sigma_{(j)}H_{(i,j)}%
^{c}=H_{(i,j)}^{c}\\
\text{F-term}  &  \text{: }\underset{j}{\sum}H_{(i,j)}\cdot H_{(i,j)}%
^{c}=0,\text{ \ \ for all }i\\
\text{D-term}  &  \text{: }\underset{j}{\sum}\left(  H_{(i,j)}\cdot
H_{(i,j)}^{\dag}-H_{(i,j)}^{c\dag}\cdot H_{(i,j)}^{c\dag}\right)
=d_{(i)}\Sigma_{(i)}\text{.}%
\end{align}
Similar considerations hold for higher order poles, as in lines
(\ref{highfirst})-(\ref{highlast}).

As we already mentioned near equation (\ref{tunecoup}), it is also natural to study the broader class of
solutions when we take generic values of the gauge couplings. Indeed, our algebraic solutions will clearly deform
smoothly (possibly at the expense of the convenient Lie algebraic structure initially used to identify the solutions) as we move to generic values of these parameters. The only subtlety is that at special tuned values of these parameters,
additional discrete symmetries may emerge, and there is a general compatibility condition between punctures which must be satisfied in
constructing models on a compact punctured Riemann surface \cite{Gaiotto:2015usa}. Since
we are concerned here with the structure of a single puncture, this subtlety plays no role in our analysis.

\subsection{Flavor Symmetries and Mass Parameters}

The symmetries of the 5D system that are not broken by boundary conditions
descend to flavor symmetries of the 4D theory localized on the puncture.\footnote{Of course, on a compact Riemann surface these
flavor symmetries will then be gauged.} For example, in the case where
we take all boundary conditions to be trivial, the resulting flavor symmetry
is at least the product of quiver gauge algebras. In principle, there can be a
further enhancement in this flavor symmetry. More generally, once we consider
non-trivial solutions to the generalized Nahm pole equations, we obtain only a
subalgebra of the quiver theory gauge algebra:%
\begin{equation}
\mathfrak{g}_{\text{flav}}\subset\underset{i}{%
%TCIMACRO{\dprod }%
%BeginExpansion
{\displaystyle\prod}
%EndExpansion
}\mathfrak{u}(Nd_{i}).
\end{equation}
Much as in other contexts, we define a \textquotedblleft complexified mass
parameter\textquotedblright\ as parameters which transform in the adjoint
representation of $\mathfrak{g}_{\text{flav}}$. One can see that the name is
appropriate by returning, for example, to equation (\ref{superpot}), in which
we can consider activating a background constant value (i.e. no singularity)
for the $Z_{(i)}$. Note that owing to the $\mathcal{N}=2$ structure of the 5D
theory, we must actually demand these mass parameters are valued in the Cartan
subalgebra of $\mathfrak{g}_{\text{flav}}$. So in other words, the mass
parameters of our theory with punctures are fully capured by the maximal
Torus:%
\begin{equation}
\{\text{Mass Parameters}\}=\mathbb{T}(\mathfrak{g}_{\text{flav}}).
\end{equation}

With these preliminaries dispensed with, let us now turn to some
representative examples of generalized Nahm pole equations.

\section{Commuting Nilpotent Matrices \label{sec:PAIRS}}

Before proceeding to the case of punctures for our $(1,0)$ theories,
let us make a few general comments on the structure of $1/4$ BPS punctures for
the $(2,0)$ theories. Indeed, all of the solutions we obtain for these $(1,0)$
theories will simply be special cases of these more general considerations.

Recall that the $1/4$ BPS\ punctures are characterized by the
equations:%
\begin{equation}
\lbrack\Sigma,Q]=Q\text{, \ \ }[\Sigma,\widetilde{Q}]=\widetilde{Q}\text{,
\ \ }[Q,\widetilde{Q}]=0\text{, \ \ }[Q,Q^{\dag}]+[\widetilde{Q}%
,\widetilde{Q}^{\dag}]=\Sigma\text{.}%
\end{equation}
As we now explain, both $Q$ and $\widetilde{Q}$ are nilpotent, so as noted in \cite{Xie:2013gma},
we get a partial characterization of solutions by enumerating pairs of commuting
nilpotent elements. A full characterization would require us to also impose
all conditions associated with $\Sigma$.

To see that $Q$ is nilpotent, first note that since $[\Sigma,Q]=Q$, we also
have $[\Sigma,Q^{l}]=l Q^{l}$ for all $l>0$. Taking the trace of each side, we
obtain the relation Tr$(Q^{l})=0$ for all $l$. This establishes the claim.

Repeating this argument for $\widetilde{Q}$, we learn that $Q$ and
$\widetilde{Q}$ are both nilpotent, and commute. As far as we are aware, the
classification of pairs of nilpotent commuting matrices is still an open
problem. There is, however, a rich connection between such pairs and
elements of the punctual Hilbert scheme for $\mathbb{C}^{2}$
(see e.g. \cite{PairsNilp} and references therein). Indeed, from
this perspective, the $1/2$ BPS\ punctures of the $(1,0)$ theories we
study are just elements of the $\Gamma$-equivariant Hilbert scheme on
$\mathbb{C}^{2}$.

Based on the fact that this classification is still an open problem, we shall
primarily focus on canonical classes of examples where the analysis is still tractable.

\section{$\mathfrak{su}(2)_{Q}$ Ansatz \label{sec:SU2}}

Perhaps the most direct analogue of the classification of punctures for
the $(2,0)$ theories are those in which we simply take the same class of
solutions, and then impose the orbifold projection constraint. For these
solutions, we find it simpler to work in the covering space basis. In this
case, we have the conditions:%
\begin{equation}
\lbrack\Sigma,Q]=Q\text{, \ \ }[Q,Q^{\dag}]=\Sigma\text{, \ \ }\widetilde{Q}%
=0\text{,}%
\end{equation}
and for each such solution we impose the orbifold projection constraint. These are the commutation relations for an $\mathfrak{su}(2)$ algebra:%
\begin{equation}
\lbrack J_{a},J_{b}]=i\varepsilon_{abc}J_{c}%
\end{equation}
in which we make the identifications:%
\begin{equation}
Q=\frac{1}{\sqrt{2}}\left(  J_{x}+iJ_{y}\right)  \text{ \ \ and \ \ }%
\Sigma=J_{z}.
\end{equation}
As standard, we also introduce the Casimir operator $J^{2}=\Sigma^2 + \{ Q, Q^\dagger \} =J_{x}^{2}+J_{y}%
^{2}+J_{z}^{2}$.

Let us now turn to the classification of solutions for the $\mathfrak{su}(2)_Q$
ansatz. As a warmup, consider the $1/2$ BPS\ punctures of A-type $(2,0)$
theories. Here, all of the data is characterized by a choice of $Q$ a
nilpotent matrix. By a suitable choice of basis, we can assume $Q$ is in
Jordan normal form, and is given by a direct of nilpotent blocks of size
$\mu_{i}\times\mu_{i}$. We can also order the $\mu_{i}$'s so that
\begin{equation}
\mu_{1}\geq\mu_{2}\geq...\geq\mu_{l},
\end{equation}
for some $l\geq1$. Since $\mu_{1}+...+\mu_{l} = N$, we label possible boundary
conditions by a choice of a partition of $N$. A convenient presentation of
this is in terms of a Young diagram. To adhere with the notation in the class
$\mathcal{S}$ literature, (rather than what is present in the representation theory
literature), we label our Young diagrams as a sequence of columns with
$\mu_{i}$ boxes in which we read the partition from left to right. Here are
examples of such Young diagrams for the partitions $[1^{N}]$,
$[N-1,1]$ and $[N]$:
\begin{align}
\lbrack1^{N}]    :~~ \underbrace{{\young(~~~\cdots~~~) }}_{\mbox{\large $N$} } ~~,~~
\Yvcentermath1 \lbrack N-1,1]  :~~  \mbox{\large $N-1$} \left\{ {\young(~,~,\vdots,~,~~) } \right.~~,~~
\Yvcentermath1 \lbrack N]  :~~  \mbox{\large $N$} \left\{ {\young(~,~,\vdots,~,~,~) } \right.
\end{align}

The partition $[1^{N}]$ defines a ``full puncture,'' while $[N-1,1]$ corresponds to a ``simple puncture,'' and  $[N]$ corresponds to an ``empty puncture.'' These three types of punctures correspond (respectively) to maximal, minimal and trivial flavor symmetries. There is an analogue of these full and simple pictures for the $(1,0)$ class $\mathcal{S}_{\Gamma}$ theories for $\Gamma = \mathbb{Z}_k$,
which was recently studied in \cite{Gaiotto:2015usa}.\footnote{The ``maximal'' (resp. ``minimal'') punctures of \cite{Gaiotto:2015usa} are expected to be the ``full'' (resp. ``simple'') punctures discussed in this paper.} We will indeed see how these specific cases fit into a much broader
class of solutions.

Along these lines, consider next the $1/2$ BPS\ punctures for our $(1,0)$ theories. First
of all, we can see that only the A-type case $\Gamma=\mathbb{Z}_{k}$ will provide non-trivial solutions when $\widetilde Q = 0$. The reason is simply that all other
groups $\Gamma$ contain generators that non-trivially rotate the doublet comprised of
$Q$ and $\widetilde{Q}$. In
the quiver basis, we have a gauge group $U(N)^{k}$, which we label as
$i=1,...,N$. The hypermultiplets are then given by links which form a
ring:\ $H_{(i+1,i)}\oplus H_{(i+1,i)}^{c}$, or simply $H_{(i)}\oplus
H_{(i)}^{c}$, where $i=N+1$ is identified with $i=1$. Our ansatz embeds in the
larger covering space as:%
\begin{equation}
Q=\left[
\begin{array}
[c]{cccc}
&  &  & H_{(N)}\\
H_{(1)} &  &  & \\
& \ddots &  & \\
&  & H_{(N-1)} &
\end{array}
\right]  \text{ \ \ and \ \ }\Sigma=\left[
\begin{array}
[c]{cccc}%
\Sigma_{(1)} &  &  & \\
& \Sigma_{(2)} &  & \\
&  & \ddots & \\
&  &  & \Sigma_{(N)}%
\end{array}
\right]  .
\end{equation}

We now proceed to classify all of the resulting punctures for this ansatz.
Again, the covering space description is most helpful. In particular, in the
basis specified above, we introduce:
\begin{equation}
\gamma=\left[
\begin{array}
[c]{ccccc}%
\omega\mathbb{I}_{N} &  &  &  & \\
& \omega^{2}\mathbb{I}_{N} &  &  & \\
&  & \ddots &  & \\
&  &  & \omega^{k-1}\mathbb{I}_{N} & \\
&  &  &  & \mathbb{I}_{N}%
\end{array}
\right]  , \label{eq:gammag}%
\end{equation}
with $\mathbb{I}_{N}$ the $N\times N$ identity matrix and $\omega$ a primitive
$k$th root of unity. This gives
\begin{equation}
\gamma\Sigma\gamma^{\dagger}=\Sigma\,,~~~~\gamma Q\gamma^{\dagger}=\omega Q. \label{eq:proj}%
\end{equation}
The solutions decompose into representations of $\mathfrak{su}(2)$:%
\begin{equation}
Q=\bigoplus_{j}r_{j}H_{(j)}%
\end{equation}
for spins $j$ and multiplicities $r_{j}$, with $\sum_{j}r_{j}(2j+1)=Nk$. The
action of $\gamma$ must be compatible with the $\mathfrak{su}(2)$ algebra.
Notice that $\gamma J^{2}\gamma^{\dagger}=J^{2}$ implies $\gamma$ preserves
the spin of the representation. Further, $\gamma\Sigma\gamma^{\dagger}=\Sigma$
so that $J^{2},\Sigma,\gamma$ all commute. We find:%
\begin{equation}
\gamma\left\vert j_{i}\text{ }m_{i}\right\rangle =a_{j_{i}m_{i}}\left\vert
j_{i}\text{ }m_{i}\right\rangle ~~~\Rightarrow~~~\gamma Q\left\vert
j_{i}\text{ }m_{i}\right\rangle =\omega a_{j_{i}m_{i}}Q\left\vert j_{i}\text{
}m_{i}\right\rangle ,~~~~~a_{j_{i}m_{i}}^{k}=1.
\end{equation}
Here, $i$ runs from $1$ to $r_{j}$. For a given $j_{i}$ we pick some a lowest
eigenvalue $a_{j_{i},-j_{i}}=a_{j_{i}}$, so that%
\begin{equation}
a_{j_{i},-j_{i}+n}=a_{j_{i}}\omega^{n}.
\end{equation}
This completely fixes the form of $\gamma$. Our solutions are thus specified
by a choice of $Nk$-dimensional representation of $\mathfrak{su}(2)$
(equivalently, a partition of $Nk$) as well as a choice of $k$th root of unity
$a_{j_{i},-j_{i}}$ for each $i$:
\begin{equation}
Q=\bigoplus_{j}\bigoplus_{n=0}^{k-1}r_{j,n}H_{j,n}, \label{eq:Asolution}%
\end{equation}
where $r_{j,n}$ labels the multiplicity of representations $H_{j,n}$ of spin
$j$ with $a_{j,-j}=\omega^{n}$, and
\begin{equation}
Nk=\sum_{j}\sum_{n=0}^{k-1}r_{j,n}(2j+1).
\end{equation}
Additionally, solutions are subject to the constraint that each $k$th root of
unity must show up precisely $N$ times as one of the $a_{j_{i},m_{i}}$.

To construct these solutions more concretely, it is helpful to work in the
\textquotedblleft Jordan basis," in which $Q$ is a nilpotent matrix with
entries along the superdiagonal,
\begin{equation}
Q=\left[
\begin{array}
[c]{ccccc}%
0 & c_{1} &  &  & \\
& 0 & c_{2} &  & \\
&  & \ddots &\ddots  & \\
&  &  & 0 & c_{Nk-1}\\
&  &  &  & 0
\end{array}
\right]  , \label{eq:qeq}%
\end{equation}
and $\Sigma$ is diagonal. Analogous to the ordinary Nahm equations for
$U(Nk)$, this shows that the solutions are labeled by partitions of $Nk$,
where each column of the partition corresponds to a decoupled Jordan block of
$Q$. For instance, the partition $[Nk]$ corresponds to the case where all the
$c_{i}$ are nonvanishing, whereas the partition $[2,1,1,...,1]$ corresponds to
the case where $c_{1}\neq0$ but the rest vanish.

In addition to the choice of partition, solutions are labeled by a choice of
$\gamma$, which we take to be diagonal. Equation (\ref{eq:gammag}) tells us the
spectrum of eigenvalues of $\gamma$, but we still have the freedom to
rearrange the eigenvalues $\lambda_{i}$ of $\gamma$ along the diagonal as we
see fit. The one additional restriction comes from (\ref{eq:proj}), which
tells us that $\lambda_{i+1}=\omega\lambda_{i}$ if $c_{i}\neq0$.

The above conditions admit a combinatorial interpretation. Solutions
to the generalized Nahm pole equations for a quiver of $k$ $U(N)$ gauge groups
are specified by Young diagrams of $Nk$ boxes. Given such a diagram, we must
fill in each box with a $k$th root of unity $\omega^{j}$ subject to the
constraints that each root of unity must appear $N$ times in the diagram, and
any box stacked on another box must have a primitive root of unity that is
$\omega$ times the root of unity in the box below it. To keep the notation
readable, we display just the exponent in each box. These are to be read
vertically from bottom to top.

Columns are indistinguishable in the sense that switching the order of two
columns of the same height does not give a new solution.
For instance, in the case of $N = 3$ and $k = 2$, (i.e., three M5-branes
at a $\mathbb{C}^2 / \mathbb{Z}_2$ singularity), the following are equivalent:
$$
\begin{array}{c}{\young(0,1,0101)}\end{array}\cong \begin{array}{c}{\young(0,1,0011)}\end{array}.
$$

There is yet another way to describe these solutions, which as we show in section
\ref{sec:PAIRS} generalizes to D- and E-type singularities. Namely, we can represent a solution by a
directed graph through a generalization of the associated affine Dynkin
diagram. As a simple case, consider the $k=4$, $N=1$ theory (One M5-brane at
a $\mathbb{C}^2 / \mathbb{Z}_4$ singularity) and the solution with partition
$$
\begin{array}{c}{\young(2,1,03)}\end{array}.
$$
Here, the labels indicate the powers of $\omega=\exp(2 \pi i / 4)$ associated with
each box. The partition tells us that the the chiral field between the gauge
groups labeled by $\omega^{0}$ and $\omega^{1}$ is turned on, as is the chiral
field between the $\omega^{1}$ and $\omega^{2}$ gauge groups. Pictorially, we
may represent this by the following directed graph between the nodes of the
$\widehat{A}_{3}$ Dynkin diagram:
$$
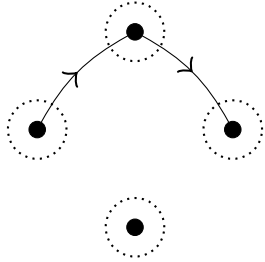
\begin{array}
[c]{c}%
\begin{tikzpicture}[scale=1.3]
\node[draw,circle,fill=black,scale=.3] (0inner) at (-1,0) {$0$};
\node[draw,circle,fill=black,scale=.3] (1inner) at (0,1) {$1$};
\node[draw,circle,fill=black,scale=.3] (2inner) at (1,0) {$2$};
\node[draw,circle,fill=black,scale=.3] (3inner) at (0,-1) {$3$};
\node[draw,circle,thick,dotted,scale=2] at (-1,0) {};
\node[draw,circle,thick,dotted,scale=2] at (0,1) {};
\node[draw,circle,thick,dotted,scale=2] at (1,0) {};
\node[draw,circle,thick,dotted,scale=2] at (0,-1) {};
\draw[->-=.5] (0inner) to [bend left = 16] (1inner);
\draw[->-=.5] (1inner) to [bend left= 16] (2inner);
\end{tikzpicture}
\end{array}
$$
This extends to theories with $N>1$. For instance, the $k=4$, $N=2$ theory has
a solution with partition
$$
\begin{array}{c}
{\young(0,3,2,13,021)}
\end{array}.
$$
This is represented by the directed graph
$$
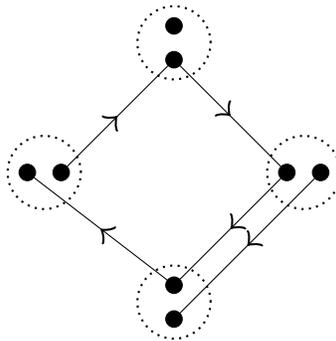
\begin{array}
[c]{c}%
\begin{tikzpicture}[scale=1.5]
\node[draw,circle,fill=black,scale=.3] (0outer) at (-1.3,0) {$0$};
\node[draw,circle,fill=black,scale=.3] (0inner) at (-1,0) {$0$}; \node[draw,circle,fill=black,scale=.3] (1outer) at (0,1.3) {$1$}; \node[draw,circle,fill=black,scale=.3] (1inner) at (0,1) {$1$};
\node[draw,circle,fill=black,scale=.3] (2outer) at (1.3,0) {$2$};
\node[draw,circle,fill=black,scale=.3] (2inner) at (1,0) {$2$};
\node[draw,circle,fill=black,scale=.3] (3outer) at (0,-1.3) {$3$};
\node[draw,circle,fill=black,scale=.3] (3inner) at (0,-1) {$3$};
\draw[->-=.5] (0inner) to(1inner);
\draw[->-=.5] (1inner) to (2inner);
\draw[->-=.5] (3inner) to (0outer);
\draw[->-=.5] (2outer) to (3outer);
\draw[->-=.5] (2inner) to (3inner);
\node[draw,circle,dotted,thick,scale=2.5] at (1.15,0){};
\node[draw,circle,dotted,thick,scale=2.5] at (0,1.15){};
\node[draw,circle,dotted,thick,scale=2.5] at (-1.15,0){};
\node[draw,circle,dotted,thick,scale=2.5] at (0,-1.15){};
\end{tikzpicture}
\end{array}
$$
Note that the vertices of this graph consist of two copies of the nodes of the
$\widehat{A}_{3}$ diagram, since $N=2$ in this case. The edges of the graph always
point clockwise around the quiver diagram. Every vertex in the graph can have
at most one incoming and one outgoing edge. In this way, every $\mathfrak{su}%
(2)$ solution for the type A quivers with anti-chirals turned off can be
represented by a directed graph, a point we return to in section
\ref{sec:GRAPH}.

\subsection{Flavor Symmetries}

Let us now turn to the continuous flavor symmetries for our puncture, i.e.,
the subalgebra of the 5D gauge symmetry which is left unbroken by our boundary
conditions. First, recall that in the case of a class $\mathcal{S}$ theory $1/2$
BPS\ puncture, the flavor symmetry associated with a partition $\{(\mu
_{1})^{r_{1}},...,(\mu_{l})^{r_{l}}\}$ is given by \cite{Chacaltana:2012zy}:%
\begin{equation}
\mathfrak{g}_{\text{flav}}=\mathfrak{s}\left[
\underset{i=1}{\overset{l}{\bigoplus}}\mathfrak{u}(r_{i})\right]  ,
\end{equation}
where $r_{i}$ is the multiplicity of a given partition. For the $1/2$ BPS
punctures of the class $\mathcal{S}_{\Gamma}$ theories with $\Gamma
=\mathbb{Z}_{k}$, we also have the data of a partition, but with a further
refinement given by the overall complex phase attached to the lowest weight
state of an irreducible representation. So, taking a further partition of
$r_{i}$ as:%
\begin{equation}
r_{i}=m_{i}^{(1)}+...+m_{i}^{(t_{i})},
\end{equation}
we get that the unbroken flavor symmetry is:%
\begin{equation}
\mathfrak{g}_{\text{flav}}=\mathfrak{s}\left[
\underset{i=1}{\overset{l}{\bigoplus}}\underset{p=1}{\overset{t_{i}}{\bigoplus}%
}\mathfrak{u}(m_{i}^{(p)}))\right]  .\label{gflavdegen}%
\end{equation}
More succinctly, we can write this as:%
\begin{equation}
\mathfrak{g}_{\text{flav}}=\mathfrak{s}\left[  \underset{m_{\text{degen}%
}}{\bigoplus}\mathfrak{u}(m_{\text{degen}})\right]  ,\label{gflavagain}%
\end{equation}
where $m_{\text{degen}}$ is the degeneracy of a given spin, with associated
$\mathbb{Z}_{k}$ charge.

\section{$\mathfrak{su}(2)_{Q}\times\mathfrak{su}(2)_{\protect\widetilde{Q}}$
Ansatz\label{sec:TWOSU2}}

To generate more examples of solutions to the generalized Nahm pole equations,
we now turn to an ansatz in which we have two independent $\mathfrak{su}(2)$
subalgebras. Returning to our system of equations,%
\begin{equation}
\lbrack\Sigma,Q]=Q\text{, \ \ }[\Sigma,\widetilde{Q}]=\widetilde{Q}\text{,
\ \ }[Q,\widetilde{Q}]=0\text{, \ \ }[Q,Q^{\dag}]+[\widetilde{Q}%
,\widetilde{Q}^{\dag}]=\Sigma\text{.}%
\end{equation}
we now impose the further condition:%
\begin{equation}
\lbrack Q,\widetilde{Q}^{\dag}]=0.
\end{equation}
Since $Q$ and $\widetilde{Q}$ are each nilpotent, this additional condition
means that our solutions will be captured by representations of $\mathfrak{su}%
(2)\times\mathfrak{su}(2)$. In this case, it is again helpful to introduce the
corresponding $\mathfrak{su}(2)$ generators $J_{a}$ and $\widetilde{J}_{a}$.

Let us now turn to the types of orbifold group projections compatible with
these conditions. Consider first the class $\mathcal{S}$ theories, where we
have a $1/4$ BPS\ punctures. Now, since we have a pair of commuting
$\mathfrak{su}(2)$s, we can decompose $\mathbb{C}^{N}$ into some choice of
irreducible representations of $\mathfrak{su}(2)\times\mathfrak{su}(2)$. For
this choice, a vector $\left\vert \Psi\right\rangle \in\mathbb{C}^{N}$ will
decompose as:%
\begin{equation}
\left\vert \Psi\right\rangle =\underset{j\text{ }m\text{ };\text{
}\widetilde{j}\text{ }\widetilde{m}\text{ };\text{ }s}{%
%TCIMACRO{\dsum }%
%BeginExpansion
{\displaystyle\sum}
%EndExpansion
}\psi_{jm;\widetilde{j}\widetilde{m}}\left\vert j\text{ }m\text{ };\text{
}\widetilde{j}\text{ }\widetilde{m}\text{ };\text{ }s\right\rangle ,
\label{genstate}%
\end{equation}
where $j$ is the spin with respect to the $\mathfrak{su}(2)$ generated by $Q$,
$m$ labels a state in this representation, and similar considerations hold for
$\widetilde{j}$ and $\widetilde{m}$ with respect to $\widetilde{Q}$. Here, $s$
is an additional index to account for the possibility that we have a
degeneracy in our decomposition, i.e., a given spin may appear more than once.

We thus need to list possible representations $(j,\widetilde{j},s)$ which
appear in such a decomposition. The choices compatible with our other
conditions are that we have a specific class of partitions:%
\begin{equation}
N=\underset{(j,\widetilde{j},s)}{%
%TCIMACRO{\dsum }%
%BeginExpansion
{\displaystyle\sum}
%EndExpansion
}\left(  2j+1\right)  (2\widetilde{j}+1),
\end{equation}
in the obvious notation.

Consider next the $1/2$ BPS\ punctures of the $(1,0)$ theories of class
$\mathcal{S}_{\Gamma}$. Here, we would like to first determine whether our
ansatz is compatible with a particular choice of $\Gamma$. In the case of A-
and D-type discrete subgroups, we will give a classification of the resulting
boundary conditions. For the E-type quivers, however, we find that there are
no non-trivial solutions. In the following subsections we step through each possibility.

\subsection{A-type $\Gamma$}

Let us now turn to the further constraints imposed by working with the
$\mathbb{Z}_{k}$ orbifold. Essentially, our task reduces to tracking the group
action of elements of $\mathbb{Z}_{k}$ on a state such as that given in
equation (\ref{genstate}). Since we
can potentially have a degeneracy for each choice of representation, we again
label states of the representation as $\left\vert j\text{ }m\text{ };\text{
}\widetilde{j}\text{ }\widetilde{m}\text{ };\text{ }s\right\rangle $. Noting that
 $\gamma, J_{z}$, $\widetilde{J}_{z}$ $J^{2},\widetilde{J}^{2}$ constitute a set of commuting normal matrices, we can without loss of generality assume that $\gamma$ has been
diagonalized by an element of $U(Nk)$ when acting on the index $s$. We then have:%
\begin{equation}
\gamma\left\vert j\text{ }m\text{ };\text{ }\widetilde{j}\text{ }%
\widetilde{m}\text{ };\text{ }s\right\rangle =a_{j\text{ }m\text{ };\text{
}\widetilde{j}\text{ }\widetilde{m}\text{ };\text{ }s\text{ }}\left\vert
j\text{ }m\text{ };\text{ }\widetilde{j}\text{ }\widetilde{m}\text{ };\text{
}s\right\rangle ,
\end{equation}
for some complex phase $a_{j\text{ }m\text{ };\text{ }\widetilde{j}\text{
}\widetilde{m}\text{ };\text{ }s\text{ }}$ subject to the condition (since
$\gamma^{k}=1$):%
\begin{equation}
\left(  a_{j\text{ }m\text{ };\text{ }\widetilde{j}\text{ }\widetilde{m}\text{
};\text{ }s\text{ }}\right)  ^{k}=1\text{.}%
\end{equation}

Next, consider the effect of acting by the raising operator (i.e., the
rescaled versions of $Q$ and $\widetilde{Q}$):%
\begin{align}
\gamma J_{+}\left\vert j\text{ }m\text{ };\text{ }\widetilde{j}\text{
}\widetilde{m}\text{ };\text{ }s\right\rangle  &  =\omega a_{j\text{ }m\text{
};\text{ }\widetilde{j}\text{ }\widetilde{m}\text{ };\text{ }s\text{ }}%
J_{+}\left\vert j\text{ }m\text{ };\text{ }\widetilde{j}\text{ }%
\widetilde{m}\text{ };\text{ }s\right\rangle ,\nonumber\\
\gamma\widetilde{J}_{+}\left\vert j\text{ }m\text{ };\text{ }\widetilde{j}%
\text{ }\widetilde{m}\text{ };\text{ }s\right\rangle  &  =\omega
^{-1}a_{j\text{ }m\text{ };\text{ }\widetilde{j}\text{ }\widetilde{m}\text{
};\text{ }s\text{ }}\widetilde{J}_{+}\left\vert j\text{ }m\text{ };\text{
}\widetilde{j}\text{ }\widetilde{m}\text{ };\text{ }s\right\rangle .
\end{align}
So as expected, for a given $j$, $\widetilde{j}$, we pick some a lowest
eigenvalue $a_{j\text{ }m\text{ };\text{ }\widetilde{j}\text{ }\widetilde{m}%
\text{ };\text{ }s\text{ }}=\alpha_{j\text{ };\text{ }\widetilde{j}\text{
};\text{ }s}$, so that further shifts in the $m$ and $\widetilde{m}$ index
obey:%
\begin{equation}
a_{j\text{ }-j+m\text{ };\text{ }\widetilde{j}\text{ }-\widetilde{j}%
+\widetilde{m}\text{ };\text{ }s}=\alpha_{j\text{ };\text{ }\widetilde{j}%
\text{ };\text{ }s}\omega^{m-\widetilde{m}}.
\end{equation}
Once again, our solutions are labeled by a collection of spins
$(j,\widetilde{j},s)$ (with possible degeneracy), as well as a choice of
$\alpha_{j\text{ };\text{ }\widetilde{j}\text{ };\text{ }s}$, subject to the
above constraints. Additionally, each $k$th root of
unity must appear $N$ times as one of the $a_{j \text{ } m;\text{ } \widetilde{j} \text{ } \widetilde{m} \text{ }; s}$.

\subsection{D-type $\Gamma$}

Consider next the case of the D-type discrete subgroups of $SU(2)$. In a quiver basis where we have $k$ simple gauge group
factors (i.e. we have a $D_k$ type singularity), the defining relations for the discrete group are:
\begin{equation}
\gamma^{2k - 4}=1\text{, \ \ }\gamma^{k - 2}=\tau^{2}\text{, \ \ }\gamma\tau\gamma=\tau. \label{Dtype}%
\end{equation}
In terms of $2\times2$ matrix representatives, we have:%
\begin{equation}
\gamma_{\text{doub}}=\left[
\begin{array}
[c]{cc}%
\omega & \\
& \omega^{-1}%
\end{array}
\right]  \text{ \ \ and \ \ }\tau_{\text{doub}}=\left[
\begin{array}
[c]{cc}%
0 & 1\\
-1 & 0
\end{array}
\right]  ,
\end{equation}
where $\omega$ is a primitive $(2k-4)$th root of unity, i.e. $\omega^{2k-4}=1$.
To obtain a solution consistent with the orbifold projection we first obtain a
solution to the A-type case, and then impose a further constraint by requiring
invariance under the action of $\tau$. Since $\gamma$ and $\tau$ generate the
group, this is sufficient to determine the algebraic structure of the solution.

We again choose to label all states as $\left\vert j\text{ }m\text{ };\text{
}\widetilde{j}\text{ }\widetilde{m}\text{ };\text{ }s\right\rangle $, where
$s$ is an index indicating the possible degeneracy with respect to a given
choice of spins. Now, we have the orbifold projection conditions:%
\begin{equation}
\tau Q\tau^{-1}=\widetilde{Q}\text{, \ \ }\tau\widetilde{Q}\tau^{-1}=-Q.
\label{eqn:Qproj}%
\end{equation}
In particular, we therefore obtain the relations:%
\begin{align}
\tau J_{z}\tau^{-1}  &  =\widetilde{J}_{z}\text{, \ \ }\tau\widetilde{J}%
_{z}\tau^{-1}=J_{z}\\
\tau J^{2}\tau^{-1}  &  =\widetilde{J}^{2}\text{, \ \ }\tau\widetilde{J}%
^{2}\tau^{-1}=J^{2}%
\end{align}

Consider, then, the matrix elements of $\tau$. Since we have $\tau
J_{z}=\widetilde{J}_{z}\tau$, we have:%
\begin{align}
\tau J_{z}\left\vert j\text{ }m\text{ };\text{ }\widetilde{j}\text{
}\widetilde{m}\text{ };\text{ }s\right\rangle  &  =\underset{j^{\prime
}m^{\prime};\widetilde{j}^{\prime}\widetilde{m}^{\prime}\text{ };\text{
}s^{\prime}}{\sum}m\tau_{jm;\widetilde{j}\widetilde{m}|j^{\prime}m^{\prime
};\widetilde{j}^{\prime}\widetilde{m}^{\prime}\text{ };\text{ }s^{\prime}%
}\left\vert j^{\prime}\text{ }m^{\prime}\text{ };\text{ }\widetilde{j}%
^{\prime}\text{ }\widetilde{m}^{\prime}\text{ };\text{ }s^{\prime
}\right\rangle \\
\tau J_{z}\left\vert j\text{ }m\text{ };\text{ }\widetilde{j}\text{
}\widetilde{m}\text{ };\text{ }s\right\rangle  &  =\underset{j^{\prime
}m^{\prime};\widetilde{j}^{\prime}\widetilde{m}^{\prime}\text{ };\text{
}s^{\prime}}{\sum}\widetilde{m}^{\prime}\tau_{jm;\widetilde{j}\widetilde{m}%
|j^{\prime}m^{\prime};\widetilde{j}^{\prime}\widetilde{m}^{\prime}\text{
};\text{ }s^{\prime}}\left\vert j^{\prime}\text{ }m^{\prime}\text{ };\text{
}\widetilde{j}^{\prime}\text{ }\widetilde{m}^{\prime}\text{ };\text{
}s^{\prime}\right\rangle .
\end{align}
We include the sum over degeneracy factors since a priori, $\tau$ may move us
between them.

From the above equations, we conclude that to have a non-zero matrix element, we need $\widetilde{m}%
^{\prime}=m$. Based on this, we learn that up to a phase, $\tau$ interchanges
the values of the spins:%
\begin{equation}
\tau\left\vert j\text{ }m\text{ };\text{ }\widetilde{j}\text{ }\widetilde{m}%
\text{ };\text{ }s\right\rangle =\underset{s^{\prime}}{\sum}c_{s;s^{\prime}%
}^{(jm;\widetilde{j}\widetilde{m})}\left\vert \widetilde{j}\text{
}\widetilde{m}\text{ };\text{ }j\text{ }m\text{ };\text{ }s^{\prime
}\right\rangle ,
\end{equation}
where to emphasize the restricted role of these phases, we have introduced a
specific collection of entries $c_{s;s^{\prime}}^{(jm;\widetilde{j}%
\widetilde{m})}$, which are the analogue of the $a_{j\text{ }m\text{ };\text{
}\widetilde{j}\text{ }\widetilde{m}\text{ };\text{ }s\text{ }}$introduced for
the A-type orbifold projection. Note that in this case, we have:%
\begin{equation}
\left(  a_{j\text{ }m\text{ };\text{ }\widetilde{j}\text{ }\widetilde{m}\text{
};\text{ }s\text{ }}\right)  ^{2k - 4}=1\text{.}%
\end{equation}

To proceed further, we ask about the relations which the $c_{s;s^{\prime}%
}^{(j\text{ }m\text{ };\text{ }\widetilde{j}\text{ }\widetilde{m})}$'s
satisfy. First, we argue that we can assume a diagonal action on the
degeneracy index $s$ and $s^{\prime}$. To see this, consider the relation
$\gamma\tau\gamma=\tau$. This does not quite yield a commutation relation.
Nevertheless, although this means we cannot simultaneously diagonalize the
operators $\gamma$ and $\tau$, it does mean that upon acting on a state of our
representation:%
\begin{align}
\gamma\tau\gamma\left\vert j\text{ }m\text{ };\text{ }\widetilde{j}\text{
}\widetilde{m}\text{ };\text{ }s\right\rangle  &  =\underset{s^{\prime}}{\sum
}a_{j\text{ }m\text{ };\text{ }\widetilde{j}\text{ }\widetilde{m}\text{
};\text{ }s\text{ }}c_{s;s^{\prime}}^{(j\text{ }m\text{ };\text{
}\widetilde{j}\text{ }\widetilde{m})}a_{\widetilde{j}\text{ }\widetilde{m}%
\text{ };\text{ }j\text{ }m\text{ };\text{ }s^{\prime}\text{ }}\left\vert
\widetilde{j}\text{ }\widetilde{m}\text{ };\text{ }j\text{ }m\text{ };\text{
}s^{\prime}\right\rangle \\
\tau\left\vert j\text{ }m\text{ };\text{ }\widetilde{j}\text{ }\widetilde{m}%
\text{ };\text{ }s\right\rangle  &  =\underset{s^{\prime}}{\sum}%
c_{s;s^{\prime}}^{(j\text{ }m\text{ };\text{ }\widetilde{j}\text{
}\widetilde{m})}\left\vert \widetilde{j}\text{ }\widetilde{m}\text{ };\text{
}j\text{ }m\text{ };\text{ }s^{\prime}\right\rangle .
\end{align}
So, we get the additional relation:%
\begin{equation}
a_{j\text{ }m\text{ };\text{ }\widetilde{j}\text{ }\widetilde{m}\text{
};\text{ }s\text{ }}a_{\widetilde{j}\text{ }\widetilde{m}\text{ };\text{
}j\text{ }m\text{ };\text{ }s^{\prime}\text{ }}=1,
\end{equation}
for all $s^{\prime}$. From this, we conclude that by a suitable change of
basis, we may assume a diagonal action for $\tau$ on our degeneracy label $s$.
We therefore adopt the notation:%
\begin{equation}
\tau\left\vert j\text{ }m\text{ };\text{ }\widetilde{j}\text{ }\widetilde{m}%
\text{ };\text{ }s\right\rangle =c_{j\text{ }m\text{ };\text{ }\widetilde{j}%
\text{ }\widetilde{m}\text{ };\text{ }s}\left\vert \widetilde{j}\text{
}\widetilde{m}\text{ };\text{ }j\text{ }m\text{ };\text{ }s\right\rangle .
\end{equation}

Returning to the relations of line (\ref{Dtype}), consider next the condition
$\gamma^{k - 2}=\tau^{2}$. Acting on states of our representation, we have:%
\begin{align}
\gamma^{k - 2}\left\vert j\text{ }m\text{ };\text{ }\widetilde{j}\text{
}\widetilde{m}\text{ };\text{ }s\right\rangle  &  =\left(  a_{j\text{ }m\text{
};\text{ }\widetilde{j}\text{ }\widetilde{m}\text{ };\text{ }s\text{ }%
}\right)  ^{k - 2}\left\vert j\text{ }m\text{ };\text{ }\widetilde{j}\text{
}\widetilde{m}\text{ };\text{ }s\right\rangle \\
\tau^{2}\left\vert j\text{ }m\text{ };\text{ }\widetilde{j}\text{
}\widetilde{m}\text{ };\text{ }s\right\rangle  &  =c_{j\text{ }m\text{
};\text{ }\widetilde{j}\text{ }\widetilde{m}\text{ };\text{ }s}%
c_{\widetilde{j}\text{ }\widetilde{m}\text{ };\text{ }j\text{ }m\text{
};\text{ }s}\left\vert j\text{ }m\text{ };\text{ }\widetilde{j}\text{
}\widetilde{m}\text{ };\text{ }s\right\rangle ,
\end{align}
from which we get the relation:%
\begin{equation}
c_{j\text{ }m\text{ };\text{ }\widetilde{j}\text{ }\widetilde{m}\text{
};\text{ }s}c_{\widetilde{j}\text{ }\widetilde{m}\text{ };\text{ }j\text{
}m\text{ };\text{ }s}=\left(  a_{j\text{ }m\text{ };\text{ }\widetilde{j}%
\text{ }\widetilde{m}\text{ };\text{ }s\text{ }}\right)  ^{k - 2}\text{.}%
\end{equation}
Accounting for the further relation:%
\begin{equation}
a_{j\text{ }-j+m\text{ };\text{ }\widetilde{j}\text{ }-\widetilde{j}%
+\widetilde{m}\text{ };\text{ }s}=\alpha_{j\text{ };\text{ }\widetilde{j}%
\text{ };\text{ }s}\omega^{m-\widetilde{m}}%
\end{equation}
with $\omega^{k - 2}=-1$, we obtain:
\begin{equation}
c_{j\text{ }-j+m\text{ };\text{ }\widetilde{j}\text{ }-\widetilde j +\widetilde{m}\text{
};\text{ }s}c_{\widetilde{j}\text{ }-\widetilde j +\widetilde{m}\text{ };\text{ }j\text{
}-j +m\text{ };\text{ }s}=(-1)^{m-\widetilde{m}}\left(  \alpha_{j\text{ };\text{
}\widetilde{j}\text{ };\text{ }s}\right)  ^{k-2}\text{.}%
\end{equation}
Accounting for the raising action of $J, \widetilde J$, we find
	\begin{align}
		c_{j ~ -j + m ~; ~\widetilde j ~ -\widetilde j +\widetilde m~;~s}	&= (-1)^{\widetilde m } \chi_{j~;~\widetilde j~;~s}
	\end{align}
which implies
	\begin{align}
		\chi_{j ~; ~ \widetilde j ~;~ s} \chi_{\widetilde j~;~ j ~; s} = \left(\alpha_{j ~ ; ~\widetilde j ~;~ s} \right)^{k-2}.
	\end{align}
We therefore also label a pair $(j,\widetilde{j},s)$ and
$(\widetilde{j},j,s)$ according to a choice of fourth roots of unity $\chi_{j\text{ };\text{
}\widetilde{j}\text{ };\text{ }s}$ and $\chi_{\widetilde j~;~ j~;~s}$.
Summarizing, then, we classify solutions to the D-type orbifold projection by
labeling representations of $\mathfrak{su}(2) \times \mathfrak{su}(2)$ with a pair of $\mathbb{Z}_{k}$ phases $\alpha_{j\text{ };\text{
}\widetilde{j}\text{ };\text{ }s},\alpha_{\widetilde j\text{ };\text{
}{j}\text{ };\text{ }s}$, and a pair of $\mathbb{Z}_{4}$ phases
$\chi_{j\text{ };\text{ }\widetilde{j}\text{ };\text{ }s}, \chi_{\widetilde j ~;~j~;~s}$ satisfying
\begin{equation}
\alpha_{j\text{ };\text{ }\widetilde{j}\text{ };\text{ }s}\alpha
_{\widetilde{j}\text{ };\text{ }j\text{ };\text{ }s}=1\text{ \ \ and \ \ }%
\chi_{j ~; ~ \widetilde j ~;~ s} \chi_{\widetilde j~;~ j ~; s} = \left(\alpha_{j ~ ; ~\widetilde j ~;~ s} \right)^{k-2}.
\end{equation}
Additionally, each $(2k -4)$th root of unity must appear $2N$ times in the eigenspectrum
of $\gamma$, while each fourth root of unity must appear $N(k-2)$ times in the eigenspectrum of $\tau$.

\subsection{E-type $\Gamma$}

Let us now demonstrate that for the E-type discrete subgroups of $\Gamma$, the
$\mathfrak{su}(2)_{Q} \times \mathfrak{su}(2)_{\widetilde{Q}}$
ansatz does not produce any non-trivial solutions.

The key point is that in contrast to the A- and D-type discrete subgroups,
here, there is always an element of the group which has $2\times2$ matrix
representative:%
\begin{equation}
\sigma_{\text{doub}}=%
\begin{bmatrix}
a & b\\
c & d
\end{bmatrix}
,~~~~~a,b,c,d\neq0.
\end{equation}
The fact that all entries are non-zero will lead to a contradiction. The
projection on the generators requires:%
\begin{align}
\sigma J_{+}\sigma^{-1} &  =aJ_{+}+b\widetilde{J}_{+}\\
\sigma\widetilde{J}_{+}\sigma^{-1} &  =cJ_{+}+d\widetilde{J}_{+}.
\end{align}
This in turn determines a conjugation rule for the $J_{z}$ and $\widetilde{J}%
_{z}$ generators:%
\begin{align}
\sigma J_{z}\sigma^{-1} &  =\left\vert a\right\vert ^{2}J_{z}+\left\vert
b\right\vert ^{2}\widetilde{J}_{z}\label{Jzone}\\
\sigma\widetilde{J}_{z}\sigma^{-1} &  =\left\vert c\right\vert ^{2}%
J_{z}+\left\vert d\right\vert ^{2}\widetilde{J}_{z}.\label{Jztwo}%
\end{align}
Now, since we also have:%
\begin{equation}
\sigma\lbrack J_{z},J_{+}]\sigma^{-1}=[\sigma J_{z}\sigma^{-1},aJ_{+}%
+b\widetilde{J}_{+}]=aJ_{+}+b\widetilde{J}_{+},
\end{equation}
we learn that:%
\begin{equation}
(\left\vert a\right\vert ^{2}-1)aJ_{+}+(\left\vert b\right\vert ^{2}%
-1)b\widetilde{J}_{+}=0
\end{equation}
Since we are assuming $J_{+}$ and $\widetilde{J}_{+}$ are linearly
independent, we learn that:%
\begin{equation}
\left\vert a\right\vert ^{2}=\left\vert b\right\vert ^{2}=1.
\end{equation}
Interchanging the roles of the $\mathfrak{su}(2)$ generators, we also obtain
the relations:%
\begin{equation}
\left\vert c\right\vert ^{2}=\left\vert d\right\vert ^{2}=1.
\end{equation}
So, returning to equations (\ref{Jzone}) and (\ref{Jztwo}), we have:%
\begin{equation}
\sigma(J_{z}+\widetilde{J}_{z})\sigma^{-1}=2(J_{z}+\widetilde{J}_{z}).
\end{equation}
But this contradicts the original orbifold projection condition:
\begin{equation}
\sigma\Sigma\sigma^{-1}=\Sigma,
\end{equation}
since $\Sigma$ is proportional to $J_{z}+\widetilde{J}_{z}$.

Summarizing, then, we conclude that to obtain non-trivial solutions for
$\Gamma$ an E-type discrete subgroup of $SU(2)$, we must seek out another ansatz.

\subsection{Flavor Symmetries}

Consider next the flavor symmetries for the $\mathfrak{su}(2)_{Q}%
\times\mathfrak{su}(2)_{\widetilde{Q}}$ ansatz. Much as in our discussion
around equation (\ref{gflavagain}), we simply need to track the degeneracy of
a given representation, i.e. the multiplicity with which it appears in our
decomposition of the puncture:%
\begin{equation}
\mathfrak{g}_{\text{flav}}=\mathfrak{s}\left[  \underset{m_{\text{degen}%
}}{\bigoplus}\mathfrak{u}(m_{\text{degen}})\right]  ,\label{mdegen}%
\end{equation}
For the A-type orbifold group projection, we just need to total up the number
of times a given pair $(j,\widetilde{j})$ appears with the same $\mathbb{Z}%
_{k}$ phase $\alpha_{j\text{ };\text{ }\widetilde{j}}$. For the D-type
orbifold projection, we seek out pairs $(j,\widetilde{j})$ and their images
under $\tau$ given by $(\widetilde{j},j)$. For each such pair, we also get a
pair of $\mathbb{Z}_{k}$ phases $\alpha_{j\text{ };\text{ }\widetilde{j}}$ and
$\alpha_{\widetilde{j}\text{ };\text{ }j}=(\alpha_{j\text{ };\text{
}\widetilde{j}})^{-1}$ with an additional $\mathbb{Z}_{4}$ phase
$\chi_{j\text{ };\text{ }\widetilde{j}}$. Again, we label the multiplicity,
and this determines the degeneracies of equation (\ref{mdegen}).

\section{$\mathfrak{su}(2)^{l}$ Directed Paths Ansatz \label{sec:GRAPH}}

In the previous sections we focused on a class of solutions which are most
transparent in the covering space basis. As we have already remarked, an
alternative but entirely equivalent way to study $1/2$ BPS\ pictures of class
$\mathcal{S}_{\Gamma}$ theories is to instead work directly with the quiver
basis. In this section we present a class of solutions which exploit this
basis to generate new solutions. Some of the solutions we arrive at have
already been encountered in the context of our $\mathfrak{su}(2)_{Q}$ and
$\mathfrak{su}(2)_{Q} \times \mathfrak{su}(2)_{\widetilde{Q}}$ solutions, though some are entirely
new. In particular, we will present a broad class of examples for all of the
E-type quivers.

The main solution generating technique we develop involves drawing a
collection of self-avoiding directed paths through the quiver. To be more
precise, we introduce some additional combinatorial data for our quiver. For
each node with gauge group $U(Nd_{i})$, we introduce $Nd_{i}$ interior vertices.
Each such vertex should be viewed as a basis vector in the vector space
$\mathbb{C}^{Nd_{i}}$. Now, for a bifundamental between $U(Nd_{i})$ and
$U(Nd_{j})$, we have a pair of linear maps:%
\begin{align}
H_{(i,j)}  & :\mathbb{C}^{Nd_{i}}\rightarrow\mathbb{C}^{Nd_{j}}\\
H_{(i,j)}^{c}  & :\mathbb{C}^{Nd_{j}}\rightarrow\mathbb{C}^{Nd_{i}}.
\end{align}
A simple collection of examples are those where we just connect one basis
vector of one node to the neighboring node. This defines a directed segment in
a link. The direction of the link tells us whether we have activated $H$ or
$H^{c}$. Note that a simple way to maintain the condition $H\cdot H^{c}=0$ is
that we take a directed path involving just the $H$'s or just the $H^{c}$'s.
Proceeding in this way, we see that we can start
to generate a directed path through the quiver. To maintain a consistent
solution, we generate a collection of paths
subject to the following rules:

\begin{enumerate}
\item Any vertex can meet at most two edges: one incoming and one outgoing.

\item Edges must connect vertices associated with adjacent nodes of the affine
Dynkin diagram.

\item Edges meeting at the same vertex must be oriented in the same direction
along the affine Dykin diagram. That is to say, an individual path can only have
$H$'s or $H^{c}$'s activated.

\item Loops are not allowed.
\end{enumerate}

The third criterion rules out paths of the form:
$$
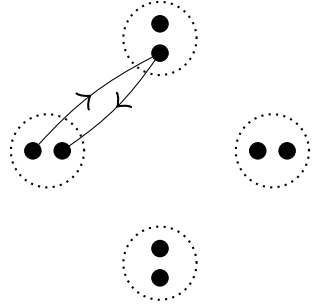
\begin{array}
[c]{c}%
\begin{tikzpicture}[scale=1.3]
\node[draw,circle,fill=black,scale=.3] (0outer) at (-1.3,0) {$0$};
\node[draw,circle,fill=black,scale=.3] (0inner) at (-1,0) {$0$}; \node[draw,circle,fill=black,scale=.3] (1outer) at (0,1.3) {$1$}; \node[draw,circle,fill=black,scale=.3] (1inner) at (0,1) {$1$};
\node[draw,circle,fill=black,scale=.3] (2outer) at (1.3,0) {$2$};
\node[draw,circle,fill=black,scale=.3] (2inner) at (1,0) {$2$};
\node[draw,circle,fill=black,scale=.3] (3outer) at (0,-1.3) {$3$};
\node[draw,circle,fill=black,scale=.3] (3inner) at (0,-1) {$3$};
\draw[->-=.5] (0outer) to [bend left=10] (1inner);
\draw[->-=.5] (1inner) to [bend left = 10] (0inner);
\node[draw,circle,dotted,thick,scale=2.5] at (1.15,0){};
\node[draw,circle,dotted,thick,scale=2.5] at (0,1.15){};
\node[draw,circle,dotted,thick,scale=2.5] at (-1.15,0){};
\node[draw,circle,dotted,thick,scale=2.5] at (0,-1.15){};
\end{tikzpicture}
\end{array}
$$
%$$
%\includegraphics[width=35mm]{badAquiver.pdf}
%$$
since this path involves activating both $H$ and $H^{c}$.
In other words, a path must continue in a fixed direction.
Here, a ``path'' is defined to be a sequence of vertices $\{v_{i}\}$
such that a directed edge points from $v_{i}$ to $v_{i+1}$. A path that bends
backwards on itself will generically violate the $[Q,\widetilde{Q}]=0$ constraint.

In the case of an A-type quiver, these conditions restrict us to the class of
$\mathfrak{su}(2)_{Q}\times\mathfrak{su}(2)_{\widetilde{Q}}$ solutions with trivial tensor
products considered previously, so it is clear that they give valid solutions.
Indeed, a chain of $m$ consecutive edges in the graph corresponds to the spin
$\frac{m}{2}$ representation of $\mathfrak{su}(2)$, with each vertex in the
chain corresponding to an eigenstate of $J_{3}$ and each edge corresponding to
an action of the raising operator $J_{+}$.

For the D- and E-type quivers, we obtain genuinely new solutions. For
instance, for the $N=1$ quiver of associated with probing a $D_5$ singularity,
one solution is represented by the directed graph
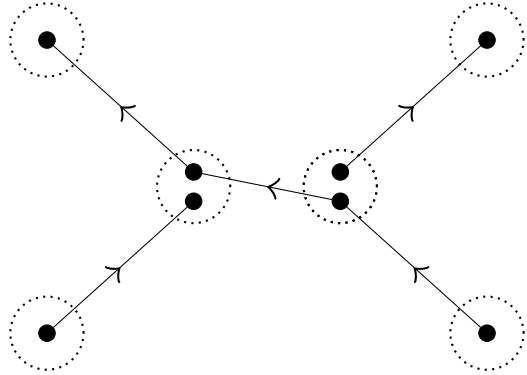
\[%
\begin{array}
[c]{c}%
\begin{tikzpicture}[scale=1.3]
\node[draw,circle,scale=.3,fill=black](1) at (0,1.5) {$1$};
\node[draw,circle,scale=.3,fill=black](2) at (0,-1.5) {$2$};
\node[draw,circle,scale=.3,fill=black](5) at (4.5,1.5) {$5$};
\node[draw,circle,scale=.3,fill=black](6) at (4.5,-1.5) {$6$};
\node[draw,circle,scale=.3,fill=black](3u) at (1.5,.15) {$3$};
\node[draw,circle,scale=.3,fill=black](3d) at (1.5,-.15) {$3$};
\node[draw,circle,scale=.3,fill=black](4u) at (3,.15) {$4$};
\node[draw,circle,scale=.3,fill=black](4d) at (3,-.15) {$4$};
\node[circle,draw,dotted,thick,scale=2.5] at (1.5,0) {};
\node[circle,draw,dotted,thick,scale=2.5] at (3,0) {};
\node[circle,draw,dotted,thick,scale=2.5] at (3,0) {};
\node[circle,draw,dotted,thick,scale=2.5] at (0,1.5) {};
\node[circle,draw,dotted,thick,scale=2.5] at (0,-1.5) {};
\node[circle,draw,dotted,thick,scale=2.5] at (4.5,-1.5) {};
\node[circle,draw,dotted,thick,scale=2.5] at (4.5,1.5) {};
\draw[->-=.5] (3u) to (1);
\draw[->-=.5] (4d) to (3u);
\draw[->-=.5] (6) to (4d);
\draw[->-=.5] (4u) to (5);
\draw[->-=.5] (2) to (3d);
\end{tikzpicture}
\end{array}
\]
%$$
%\includegraphics[width=50mm]{D5quiver.pdf}
%$$
There are two vertices for each of the middle nodes because their Dynkin index
is 2, whereas each of the outer nodes has Dynkin index 1.

As another example, consider the $N=1$ theory for the quiver coming from probing an $E_{6}$ singularity. One
solution of this form is the following:
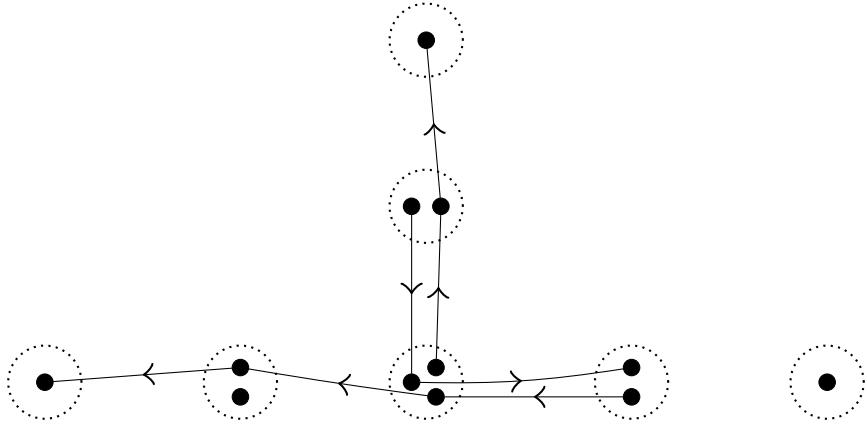
\[%
\begin{array}
[c]{c}%
\begin{tikzpicture}[scale=1.3]
\node[draw,circle,scale=.3,fill=black] (1) at (0,0){$1$};
\node[draw,circle,scale=.3,fill=black] (2u) at (2,.15){$2$};
\node[draw,circle,scale=.3,fill=black] (2d) at (2,-.15){$2$};
\node[draw,circle,scale=.3,fill=black] (3u) at (4,.15){$3$};
\node[draw,circle,scale=.3,fill=black] (3d) at (4,-.15){$3$};
\node[draw,circle,scale=.3,fill=black] (3m) at (3.75,0) {$3$};
\node[draw,circle,scale=.3,fill=black] (6u) at (6,.15){$6$};
\node[draw,circle,scale=.3,fill=black] (6d) at (6,-.15){$6$};
\node[draw,circle,scale=.3,fill=black] (7) at (8,0){$7$};
\node[draw,circle,scale=.3,fill=black] (4r) at (4.05,1.8){$4$};
\node[draw,circle,scale=.3,fill=black] (4l) at (3.75,1.8){$4$};
\node[draw,circle,scale=.3,fill=black] (5) at (3.9,3.5){$5$};
\node[circle,draw,dotted,scale=2.5,thick] at (2,0) {};
\node[circle,draw,dotted,scale=2.5,thick] at (6,0) {};
\node[circle,draw,dotted,scale=2.5,thick] at (0,0) {};
\node[circle,draw,dotted,scale=2.5,thick] at (8,0) {};
\node[circle,draw,dotted,scale=2.5,thick] at (3.9,0) {};
\node[circle,draw,dotted,scale=2.5,thick] at (3.9,1.8) {};
\node[circle,draw,dotted,scale=2.5,thick] at (3.9,3.5) {};
\draw[->-=.5] (2u) to (1);
\draw[->-=.5] (3d) to [bend left = 1] (2u);
\draw[->-=.5] (4l) to (3m);
\draw[->-=.5] (3u) to (4r);
\draw[->-=.5] (4r) to (5);
\draw[->-=.5] (3m) to [bend right=5] (6u);
\draw[->-=.5] (6d) to (3d);
\end{tikzpicture}
\end{array}
\]
%$$
%\includegraphics[width=70mm]{e6quiver.pdf}
%$$
Here, the multiplicity of
vertices associated with each node corresponds to the Dynkin number of the node.

Given this simple structure, it is natural to ask about the flavor symmetry
left unbroken by a choice of directed path. It is given by a product:%
\begin{equation}
\mathfrak{g}_{F}=\mathfrak{s}\left[  \bigoplus_{i}\mathfrak{u}(n_{i})\right]  .
\end{equation}
Here, $i$ runs over the distinct path types in the quiver, and $n_{i}$ is the
number of paths of each type, where two paths $\{v_{i}\},\{v_{i}^{\prime}\}$
are said to be of the same \textquotedblleft type\textquotedblright\ if
$v_{i}$ and $v_{i}^{\prime}$ are vertices associated with the same Dynkin node
for all $i$. Clearly, paths of the same type must be the same length, and path
type defines an equivalence relation between paths. An isolated vertex is
considered to be a path of length $1$. Intuitively, we can think of each
$\mathfrak{u}(n_{i})$ summand in the symmetry algebra as rotating the $n_{i}$
paths of identical type $i$ into each other.

As an example, consider the following directed graph solution for the A-type quiver with gauge group $U(3)^4$:
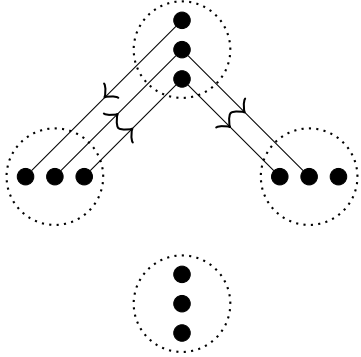
\[%
\begin{array}
[c]{c}%
\begin{tikzpicture}[scale=1.3]
\node[draw,circle,fill=black,scale=.3] (0outer+) at (-1.6,0) {$0$};
\node[draw,circle,fill=black,scale=.3] (0outer) at (-1.3,0) {$0$};
\node[draw,circle,fill=black,scale=.3] (0inner) at (-1,0) {$0$};
\node[draw,circle,fill=black,scale=.3] (1outer) at (0,1.3) {$1$};
\node[draw,circle,fill=black,scale=.3] (1inner) at (0,1) {$1$};
\node[draw,circle,fill=black,scale=.3] (1outer+) at (0,1.6) {$1$};
\node[draw,circle,fill=black,scale=.3] (2outer+) at (1.6,0) {$2$};
\node[draw,circle,fill=black,scale=.3] (2outer) at (1.3,0) {$2$};
\node[draw,circle,fill=black,scale=.3] (2inner) at (1,0) {$2$};
\node[draw,circle,fill=black,scale=.3] (3outer) at (0,-1.3) {$3$};
\node[draw,circle,fill=black,scale=.3] (3inner) at (0,-1) {$3$};
\node[draw,circle,fill=black,scale=.3] (3outer+) at (0,-1.6) {$3$};
\node[draw,circle,dotted,thick,scale=3.3] at (0,1.3){};
\node[draw,circle,dotted,thick,scale=3.3] at (0,-1.3){};
\node[draw,circle,dotted,thick,scale=3.3] at (1.3,0){};
\node[draw,circle,dotted,thick,scale=3.3] at (-1.3,0){};
\draw[->-=.5] (1inner) to (2inner);
\draw[->-=.5] (0inner) to (1inner);
\draw[->-=.5] (0outer) to (1outer);
\draw[->-=.5] (1outer) to (2outer);
\draw[->-=.5] (1outer+) to (0outer+);
\end{tikzpicture}
\end{array}
\]
%$$
%\includegraphics[width=38mm]{Directed_Flavor.pdf}
%$$
In this diagram, there are two paths of length three that begin at the left cluster of (three) vertices and terminate on the right cluster. These two paths are thus of
the same type and so contribute $\mathfrak{u}(2)$ to the flavor symmetry.
There is one path of length two (contributing $\mathfrak{u}(1)$) and four
paths of length 1, three of which are of the same type (corresponding to the
three vertices of the bottom node, contributing $\mathfrak{u}(3)$) and one of
which is of a different type (corresponding to a vertex in the right node, contributing $\mathfrak{u}(1)$).
Thus, the overall flavor symmetry is:%
\begin{equation}
\mathfrak{s}\left[  \mathfrak{u}(3)\oplus\mathfrak{u}(2)\oplus\mathfrak{u}%
(1)\oplus\mathfrak{u}(1)\right]  .
\end{equation}

\section{Conclusions \label{sec:CONC}}

Compactifications of higher-dimensional CFTs provide a general template for
realizing a rich class of lower-dimensional quantum field theories.\ In this
paper we have given a general characterization of $1/2$ BPS regular punctures
of $(1,0)$ SCFTs defined by a stack of M5-branes probing an ADE\ singularity:  class $\mathcal{S}_{\Gamma}$ theories. By compactifying these 6D
theories on a cylinder, we have shown how boundary conditions that preserve
four real supercharges reduce to a generalization of the Nahm pole equations.
We have also presented some canonical examples of solutions to these
equations, illustrating how the notion of nilpotent orbits for $1/2$
BPS\ punctures of the class $\mathcal{S}$ theories generalizes for $1/4$
BPS\ punctures of class $\mathcal{S}$ and $1/2$ BPS\ punctures of class
$\mathcal{S}_{\Gamma}$ to pairs of commuting nilpotent elements subject to
additional constraints arising from an orbifold projection. In the remainder
of this section we discuss some open areas of investigation for future work.

Our primary emphasis in this work has centered on giving various methods for
generating solutions to the generalized Nahm pole equations. Since the task of
classifying pairs of commuting nilpotent matrices is still an open problem, we
expect that a full classification of such punctures will likely be more
challenging to achieve. Nevertheless, for low rank theories, i.e.
theories defined by a single M5-brane probing an ADE singularity, we expect
that a classification should be possible.

One of the motivations for this work was to better understand the 4D theories
generated by compactification of the $(1,0)$ 6D\ SCFTs. Since class
$\mathcal{S}_{\Gamma}$ theories form the basic building blocks for more general 6D
SCFTs, it would be interesting to extend our analysis to all 6D\ SCFTs.

With the structure of punctures in place, the next step would be to
understand in more detail the structure of the resulting 4D theories. In
particular, it would be interesting to track the contributions such punctures
make to various quantities of interest in 4D such as the anomaly polynomial and
(if the compactified theory is an interacting SCFT), the superconformal index.

Finally, it is tempting to contemplate the extension of our analysis to
lower-dimensional compactifications.
Developing the analogue of the generalized Nahm pole equations in these
cases as well would provide another connection between higher-dimensional
SCFTs and their lower-dimensional descendants.

\section*{Acknowledgements}

We thank T. Iarrobino, S. Razamat and Y. Tachikawa for helpful discussions.
We also thank the Simons Center for Geometry and Physics 2016 summer workshop for hospitality during the
completion of this work. JJH also thanks the theory groups at Columbia
University, the ITS at the CUNY\ graduate center, and the CCPP at NYU for
hospitality during this work. The work of JJH is supported by NSF CAREER grant
PHY-1452037. JJH also acknowledges support from the Bahnson Fund at UNC Chapel
Hill as well as the R.~J. Reynolds Industries, Inc. Junior Faculty Development
Award from the Office of the Executive Vice Chancellor and Provost at UNC
Chapel Hill. The work of PJ, TR\ and CV is supported by NSF grant PHY-1067976.
TR is also supported by the NSF GRF under DGE-1144152.

%%%%%%%%%%%%%%%%%%%%%%%%%%%%%%%%%%%%%%%%%%%%%%%%%%%%%

\appendix

\section{Further Examples} \label{app:FURTHER}

In this Appendix we present some additional examples of solutions to the generalized Nahm pole equations
for systems with a small number $N$ of M5-branes, and for $\Gamma$ of low order. First, we give a complete classification
for solutions to the $U(3)^2$ quiver generated by $N = 3$ M5-branes probing an $A_1$ singularity.
We follow this with an example of a single M5-brane (i.e., $N = 1$) probing a $D_{4}$ singularity.

\subsection{Three M5-branes Probing an $A_1$ Singularity}

As an example, we now classify solutions to the generalized Nahm equations for the $U(3)^2$ quiver with anti-chirals turned off, $\widetilde Q = 0$.  We further work out one particular solution in detail and show how it transforms to the original basis of (\ref{eq:qeq}).  We then consider the solutions of the above form with anti-chirals turned on.

With anti-chirals turned off, there are 31 solutions to the generalized Nahm equations, given by the following Young diagrams and their inverses obtained by swapping labels, $0 \leftrightarrow 1$:
\begin{equation}
\begin{array}{ccccc}
{\young(1,0,1,0,1,0)}&{\young(0,1,0,1,01)}  & {\young(0,1,01,10)}& {\young(0,1,00,11)}&{\young(0,1,0,110)} \\\\
  {\young(10,01,10)}&{\young(11,00,11)}&{\young(0,11,001)} &{\young(1,01,100)}
& {\young(1,0,1100)}\\\\  {\young(0,1,0110)} & {\young(101,010)} &{\young(111,000)} &{\young(10,0110)} &
 {\young(00,1110)}  \\\\ {\young(1,01100)} & {\young(101010)}
\end{array}
\label{eq:solns}
\end{equation}
Note that there are only 31 solutions, rather than 34, because the Young diagrams
$$
\begin{array}{ccc}
 {\young(10,01,10)} & {\young(10,0110)} &    {\young(101010)}
\end{array}
$$
 are equivalent to their inverses.

For a concrete example of describing a solution in terms of this partition data, consider the solution labeled by the Young diagram
$$
{\young(0,1,01,10)}.
$$
In directed graph notation, this is given by
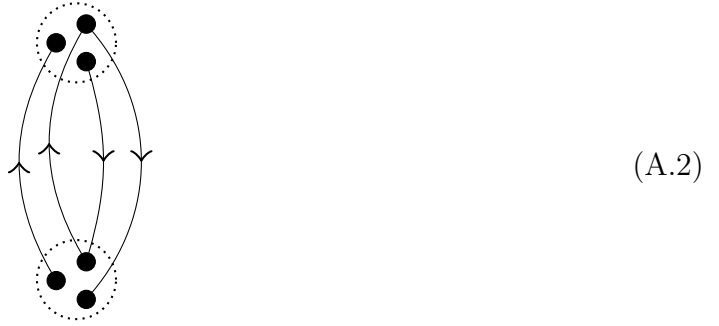
\begin{align}
	\begin{array}{c}
		\begin{tikzpicture}
			\node[draw,circle,scale=.3,fill=black] (-u) at (0,1) {$-$};
			\node[draw,circle,scale=.3,fill=black] (-d) at (0,.5) {$-$};
			\node[draw,circle,scale=.3,fill=black] (-m) at (-.4,.75) {$-$};
			\node[draw,circle,scale=.3,fill=black,yshift=-300] (+u) at (0,1) {$+$};
			\node[draw,circle,scale=.3,fill=black,yshift=-300] (+d) at (0,.5) {$+$};
			\node[draw,circle,scale=.3,fill=black,yshift=-300] (+m) at (-.4,.75) {$+$};
			\draw[->-=.5] (+m) to [bend left=30] (-m);
			\draw[->-=.5] (-u) to [bend left = 40] (+d);
			\draw[->-=.5] (-d) to [bend left=16] (+u);
			\draw[->-=.5] (+u) to [bend left = 30] (-u);
			\node[draw,circle,dotted,thick,scale=2.7] at (-.13,.75){};
			\node[draw,circle,dotted,thick,scale=2.7] at (-.13,-2.4){};
		\end{tikzpicture}
	\end{array}
\end{align}
%$$
%\includegraphics[width=25mm]{Quiver_Example.pdf}
%$$
In the Jordan basis, the corresponding $Q$ takes the form
\begin{equation} Q = \left[
\begin{array}{cccccc}
   0  & c_{1} & && &\\
 & 0& c_{2} &&&  \\
 &&  0& c_3&&\\
& &  &0 & 0& \\
&&&&0&c_5 \\
&&&&&0
  \end{array} \right]
\end{equation}
We can fix the magnitude of each $c_i$ by using the relation $[Q,Q^\dag] = \Sigma$.
The result is,
\begin{equation}
|c_1|^2 =|c_3|^2=3/2 \,,~~~~|c_2|^2=2 \,,~~~~ |c_5|^2 = 1/2.
\end{equation}
The phases can be eliminated using the $U(3)^2$ gauge symmetry.  In this basis, $\gamma$ takes the form,
\begin{equation} \gamma_g  = \left[
\begin{array}{cccccc}
   1  & & && &\\
 & -1&  &&&  \\
 &&  1& &&\\
& &  &-1 & & \\
&&&&1& \\
&&&&&-1
  \end{array} \right]
\end{equation}
To get solutions for the original fields $H_i$, $\Sigma_i$, we need to transform back to the basis in which $\gamma$ takes the form in \ref{eq:gammag}.  This is accomplished by simply permuting the eigenvalues.  Performing the same basis transformation on $Q$ gives,
\begin{equation} Q = \left[
\begin{array}{cccccc}
   0  & 0 & 0&0&0 &0\\
 0& 0& 0 &0&0& \sqrt{2}  \\
 0& 0 &  0& 0 & 0&0 \\
0 &\sqrt{3/2} & 0&0 & 0& 0\\
0&0&\sqrt{1/2}&0&0&0 \\
\sqrt{3/2}&0&0&0&0&0
  \end{array} \right]
\end{equation}
From this, we find
\begin{equation}
H_{(1,2)}  = \left[
\begin{array}{ccc}
   0  & 0 & 0\\
 0& 0& \sqrt{2}   \\
 0& 0 &  0
  \end{array} \right] \,,~~~~ H_{(2,1)}  = \left[
\begin{array}{ccc}
0 &\sqrt{3/2} & 0\\
0&0&\sqrt{1/2} \\
\sqrt{3/2}&0&0
  \end{array} \right].
\end{equation}
And,
\begin{equation}
\Sigma_1  = \left[
\begin{array}{ccc}
   -3/2  & 0 & 0\\
 0& 1/2& 0   \\
 0& 0 &  -1/2
  \end{array} \right] \,,~~~~\Sigma_2  = \left[
\begin{array}{ccc}
3/2 &0 & 0\\
0&1/2&0 \\
0&0&-1/2
  \end{array} \right].
\end{equation}
As can be checked, these matrices satisfy the generalized Nahm equations in the quiver basis.

Finally, let us consider turning on anti-chirals $H^c_{(i,j)} \neq 0$ so that the algebra splits into a decoupled $\mathfrak{su}(2) \times \mathfrak{su}(2)$.  We may then label our solutions by two decoupled Young diagrams, one for the chirals and one for the anti-chirals.  For the first six Young diagrams in (\ref{eq:solns}), there are only two that permit non-trivial solutions with anti-chirals.  Namely, we may have
\begin{align}
\begin{split}
&\left(\tfrac{3}{2} ,0\right)_{0} \oplus \left(0,\tfrac{1}{2} \right)_0  \\
&\left(\tfrac{3}{2},0\right)_{0} \oplus \left(0,\tfrac{1}{2}\right)_1 \\
&\left(1,\tfrac{1}{2}\right)_{0}
\end{split}
\end{align}
and their inverses, obtained by the interchange $0 \leftrightarrow 1$.  Here, $(j,\widetilde j )_p$ indicates the tensor product of the spin $j$ and spin $\widetilde j$ representations of $\mathfrak{su}(2)$, and the subscript labels the value of $a_{j\text{ }-j\text{ };\text{ }\widetilde{j}\text{
}-\widetilde{j}} = \omega^p$ for the representation, where $\omega$ is a primitive $k$th root of unity.

For the remaining Young diagrams in (\ref{eq:solns}), we have the non-trivial anti-chiral solutions
\begin{align}
\begin{split}
&\left(1,0\right) \oplus \left(0,1\right) \\
&\left(1,0\right) \oplus \left(0,\tfrac{1}{2}\right) \oplus \left(0,0\right) \\
&\left(\tfrac{1}{2},1\right) \\
&\left(\tfrac{1}{2},\tfrac{1}{2}\right) \oplus \left(\tfrac{1}{2},0\right)  \\
&\left(\tfrac{1}{2},0\right) \oplus \left(\tfrac{1}{2},0\right) \oplus \left(0,\tfrac{1}{2}\right)  \\
&\left(\tfrac{1}{2},\tfrac{1}{2}\right) \oplus \left(0,\tfrac{1}{2}\right)   \\
&\left(\tfrac{1}{2},\tfrac{1}{2}\right) \oplus (0,0) \oplus (0,0) \\
&\left(\tfrac{1}{2},0\right) \oplus \left(0,\tfrac{3}{2}\right)  \\
&\left(\tfrac{1}{2},0\right) \oplus \left(0,1\right) \oplus (0,0) \\
&\left(\tfrac{1}{2},0\right) \oplus \left(0,\tfrac{1}{2}\right) \oplus \left(0,\tfrac{1}{2}\right)  \\
&\left(\tfrac{1}{2},0\right) \oplus \left(0,\tfrac{1}{2}\right) \oplus (0,0) \oplus (0,0) .
\end{split}
\end{align}
For the sake of brevity, we have suppressed the subscript labels $a_{j\text{ }-j\text{ };\text{ }\widetilde{j}\text{
}-\widetilde{j}}$ for the representation, which may be filled in according to the usual rules.  Finally, for $\mu = \begin{array}{c} {\young(010101)} \end{array} $, the chiral fields are completely turned off, and the anti-chirals solutions are in one-to-one correspondence with the Young diagrams of (\ref{eq:solns}).

\subsection{One M5-brane Probing a $D_4$ Singularity}
	\label{eqn:D4}
			In this case we have a five node quiver, with a single copy of the defining representation of $\Gamma$ for the middle node. Here, we assume that all gauge couplings for the quiver are equal. There are four one-dimensional representations for the satellite nodes. The regular representation $V_{|\Gamma|} \cong \mathbb{C}^8$ decomposes as:
				\begin{align}
					V_{|\Gamma|} = V^{+}_{+} \oplus V_{+}^{-} \oplus V_1 \oplus V_1  \oplus V_{-}^{+} \oplus V_{-}^{-} , ~~~~~V_{\pm{}}^{{\pm{}}^{\prime}} \cong \mathbb C, ~~~~~ V_{1} \cong \mathbb C^2.
				\end{align}
			We are free to choose a basis in which each matrix is explicitly a direct sum of the irreducible representations described above:
				\begin{align}
					\gamma &=
							1 \oplus 1 \oplus \begin{bmatrix} i & \\ & -i \end{bmatrix} \oplus  \begin{bmatrix} i & \\ & -i \end{bmatrix}  \oplus -1 \oplus -1  \\
							\tau &= 1 \oplus -1 \oplus \begin{bmatrix} & 1 \\ -1 & \end{bmatrix} \oplus  \begin{bmatrix} & 1 \\ -1 & \end{bmatrix} \oplus -1 \oplus 1.
				\end{align}
			Imposing the orbifold projection on $\Sigma$, we learn
				\begin{align}
					\Sigma = \Sigma^{+}_{+} \oplus \Sigma_{+}^- \oplus \begin{bmatrix} \Sigma_{11} \mathbb{I}_2 & \Sigma_{12} \mathbb{I}_2 \\ \Sigma_{21} \mathbb{I}_2 & \Sigma_{22} \mathbb{I}_2 \end{bmatrix} \oplus \Sigma_{-}^{+} \oplus \Sigma_{-}^-,~~~~~ \mathbb{I}_2 = \begin{bmatrix} 1 & 0 \\ 0 & 1 \end{bmatrix}.
				\end{align}
			We should think of the middle $4 \times 4$ block of $\Sigma$ as transforming in the adjoint representation of the group $U(2)$,
				\begin{align}
					\Sigma_1 \equiv \begin{bmatrix} \Sigma_{11} \mathbb{I}_2 & \Sigma_{12} \mathbb{I}_2 \\ \Sigma_{21} \mathbb{I}_2 & \Sigma_{22} \mathbb{I}_2 \end{bmatrix} \cong \begin{bmatrix} \Sigma_{11}  & \Sigma_{12}  \\ \Sigma_{21}  & \Sigma_{22}  \end{bmatrix}.
				\end{align} 	
			Next, imposing the orbifold projection on $Q, \widetilde{Q}$, we get
				\begin{align}
					Q &=\left[
\begin{array}{cc|cccc|cc}
 0 & 0 & 0 & a_1 & 0 & a_2 & 0 & 0 \\
 0 & 0 & 0 & b_1 & 0 & b_2 & 0 & 0 \\\hline
 c_1 & d_1 & 0 & 0 & 0 & 0 & 0 & 0 \\
 0 & 0 & 0 & 0 & 0 & 0 & e_1& f_1 \\
 c_2 & d_2 & 0 & 0 & 0 & 0 & 0 & 0 \\
 0 & 0 & 0 & 0 & 0 & 0 & e_2  & f_2 \\\hline
 0 & 0 & g_1 & 0 & g_2 & 0 & 0 & 0 \\
 0 & 0 & h_1 & 0 & h_2 & 0 & 0 & 0 \\
\end{array}
\right]\\
\widetilde Q&=\left[
\begin{array}{cc|cccc|cc}
 0 & 0 &a_1 & 0 & a_2 & 0 & 0 & 0 \\
 0 & 0 & -b_1 & 0 & -b_2 & 0 & 0 & 0 \\\hline
 0 & 0 & 0 & 0 & 0 & 0 & -e_1 & f_1 \\
 -c_1 & d_1 & 0 & 0 & 0 & 0 & 0 & 0 \\
 0 & 0 & 0 & 0 & 0 & 0 & -e_2 & f_2 \\
 -c_2 & d_2 & 0 & 0 & 0 & 0 & 0 & 0 \\\hline
 0 & 0 & 0 & g_1 & 0 & g_2 & 0 & 0 \\
 0 & 0 & 0 & -h_1 & 0 & -h_2 & 0 & 0 \\
\end{array}
\right].
				\end{align}
		Of particular interest will be the row and column vectors
			\begin{align}
				\vec a^t &= \begin{bmatrix} a_1 & a_2 \end{bmatrix},~~\vec b^t = \begin{bmatrix} b_1 &  b_2 \end{bmatrix},~~\vec g^t = \begin{bmatrix} g_1 & g_2 \end{bmatrix},~~\vec h^t = \begin{bmatrix} h_1 & h_2 \end{bmatrix} \\
				\vec c &= \begin{bmatrix} c_1 \\ c_2\end{bmatrix},~~\vec d = \begin{bmatrix} d_1 \\ d_2 \end{bmatrix} , ~~ \vec e = \begin{bmatrix} e_1 \\  e_2 \end{bmatrix},~~ \vec f = \begin{bmatrix} f_1 \\ f_2 \end{bmatrix}.
			\end{align}
The above row and column vectors are precisely the bifundamental maps, as can be seen by acting with $Q, \widetilde Q$ on an arbitrary vector $\vec v \in \mathbb C^8$, being careful to restrict to a 2d diagonal subspace of the 4d space $V_1 \oplus V_1$, namely $V_1 \oplus V_1 \hookrightarrow  (V_1)_{\text{diag}}  \cong \mathbb C^2$.

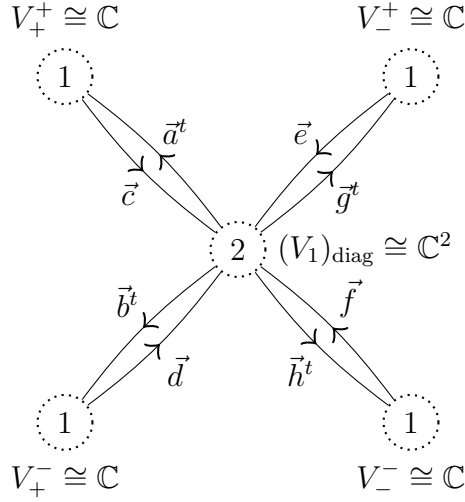
\begin{figure}
		\begin{center}
			\begin{tikzpicture}[scale=2.3]
				\node[draw,dotted,thick,circle,label={above:$V_+^+ \cong \mathbb C$}](1) at (0,0) {$1$};
				\node[draw,dotted,thick,circle,label={above:$V^+_- \cong \mathbb C$}](2) at (2,0) {$1$};
				\node[draw,dotted,thick,circle,label={below:$V^-_+ \cong \mathbb C$}](3) at (0,-2) {$1$};
				\node[draw,dotted,thick,circle,label={below:$V^-_- \cong \mathbb C $}](4) at (2,-2) {$1$};
				\node[draw,dotted,thick,circle,label={right:$(V_1)_{\text{diag}} \cong \mathbb C^2$}](5) at (1,-1) {$2$};
				\draw[->-=.5] (1) to[bend right = 8] node[below,midway]{$\vec c~~~$} (5);
				\draw[->-=.5] (5) to[bend right = 8 ] node[above,midway]{$~~~\vec a^t$}(1);
				\draw[->-=.5] (2) to[bend right = 8] node[above,midway]{$\vec e~~~$} (5);
				\draw[->-=.5] (5) to[bend right = 8] node[below,midway]{$~~~\vec g^t$} (2);
				\draw[->-=.5] (5) to[bend right = 8] node[below,midway]{$\vec h^t~~~$}(4);
				\draw[->-=.5] (4) to[bend right = 8] node[above,midway]{$~~~\vec f$} (5);
				\draw[->-=.5] (3) to[bend right = 8] node[below,midway]{$~~~\vec d$}(5);
				\draw[->-=.5] (5) to[bend right = 8] node[above,midway]{$\vec b^t~~~$}(3);
			\end{tikzpicture}
		\end{center}
		\caption{$\widehat{D_4}$ quiver diagram. Each link is labeled by the bifundamental field mapping between representation spaces. The four row vectors $\vec a^t, \vec b^t, \vec g^t, \vec h^t$ are in the representation $(2, \overline{1})$ of U(2) $\times$ U(1), while the four column vectors $\vec c, \vec d, \vec e, \vec f$ are in the representation $(1,\overline 2)$ of U(1) $\times$ U(2). Since all of the bifundamental hypermultiplets are organized in the $8 \times 8$ matrix $Q$, the matrix $\widetilde Q$ provides redundant information.}
				\label{fig:D4}
	\end{figure}

Similarly, one can identify the bifundamental maps associated to $\widetilde Q$. It is evident from the direct sum decomposition of $\tau$ that the bifundamentals of $\widetilde Q$ are related to those of $Q$ by a trivial interchanging of the two (identified) summands of $V_1 = \mathbb C \oplus \mathbb C$, where the second summand picks up a sign in the process. Therefore, it is only necessary to study $Q$ to correctly identify the bifundamental maps. The action of the bifundamental maps on the representation spaces is specified completely by the quiver diagram in Figure \ref{fig:D4}.
Using this interpretation of the bifundamental maps, one can read off the $1/2$ BPS equations directly from the matrix equations constraining $\Sigma, Q$, and $\widetilde Q$. The F-term equations are
	\begin{align}
		\Sigma_+^{+} \vec a^t - \vec a^t \Sigma_1 &= \vec a^t\\
		\Sigma^+_- \vec b^t - \vec b^t \Sigma_1 & = \vec b^t\\
		\Sigma^-_+ \vec g^t - \vec g^t \Sigma_1 &= \vec g^t \\
		\Sigma^-_- \vec h^t - \vec h^t \Sigma_1 &= \vec h^t\\
		\Sigma_1 \vec c - \vec c \Sigma_+^+ &= \vec c\\
		\Sigma_1 \vec d - \vec d \Sigma_-^+ &= \vec d\\
		\Sigma_1 \vec e - \vec e \Sigma_+^- &= \vec e\\
		\Sigma_1 \vec f - \vec f \Sigma_-^- &= \vec f,
	\end{align}
and the D-term equations are:
\begin{align}
		  \Sigma^+_+ &=2( | \vec a |^2  -  | \vec c |^2 )\\
		  \Sigma_-^+ &= 2 ( | \vec b |^2  - | \vec d |^2)  \\
		  \Sigma^-_+ & =2 (    | \vec g|^2-| \vec e |^2 )  \\
		 \Sigma_-^- &= 2 (   | \vec h |^2-|\vec f |^2 )
\end{align}
as well as:
\begin{equation}
\Sigma_{1} =(\vec c \otimes \vec c^* + \vec d \otimes \vec d^* + \vec e \otimes \vec e^* + \vec f \otimes \vec f^* )- (\vec a^* \otimes \vec a + \vec b^* \otimes \vec b + \vec g^* \otimes \vec g + \vec h^* \otimes \vec h)
\end{equation}
Furthermore, the equations corresponding to the commutation condition $[Q,\widetilde Q]= 0$ are
	\begin{align}
		0 &= \vec a \cdot \vec c = \vec b \cdot \vec d = \vec g \cdot \vec e = \vec h \cdot \vec f \\
		0 &= (\vec c \otimes \vec a + \vec e \otimes \vec g ) - (\vec d \otimes \vec b + \vec f \otimes \vec h ).
	\end{align}	

We now use the structure of representations of the algebra $\mathfrak{su}(2) \times \mathfrak{su}(2)$ to construct some simple examples of solutions to the vacuum equations for a type D quiver gauge theory. For our first example, we consider the representation
	\begin{align}
		R = \left( \tfrac{3}{2} , 0 \right) \oplus \left(0 ,\tfrac{3}{2}\right),
	\end{align}
which implies that in terms of the generators $J_a$ and $\widetilde{J}_a$ of the two $\mathfrak{su}(2)$s:
	\begin{align}
		J_+ = J_{+,\frac{3}{2} } \oplus J_{+,0} ,~~~~~J = J_{\frac{3}{2}} \oplus J_{0} ,~~~~~ \widetilde J_+ = \widetilde J_{+,0} \oplus \widetilde J_{+,\frac{3}{2}},~~~~~ \widetilde J = \widetilde J_0 \oplus \widetilde J_{\frac{3}{2}}.
			\end{align}
Given this kind of representation, the phases defining the representatives of the $\Gamma$ generators simplify to
	\begin{align}
		a_{\frac{3}{2} ~ m - \frac{3}{2}~ ;~ 0 ~0 }=  \omega^m \alpha_{-\frac{3}{2}~;~0},~~~~~c_{\frac{3}{2} ~ m - \frac{3}{2} ~;~ 0 ~0 }  = \chi_{-\frac{3}{2}~;~0}
	\end{align}
and similarly for the associated to the irrep $\left(0,\tfrac{3}{2}\right)$ (note, however, the minus sign appearing in definition of $c_{0~ 0~;~\frac{3}{2} ~ \widetilde m-\frac{3}{2}}$. Keeping in mind the fact that $j = j^\prime = \frac{3}{2}$, the consistency conditions then become
	\begin{align}
		\alpha_{-\frac{3}{2}~;~0}^2 =\alpha_{0~;~-\frac{3}{2}}^2 =  \chi_{-\frac{3}{2}~;~0}  \chi_{0~;~-\frac{3}{2}}, ~~~~~ \alpha_{-\frac{3}{2}~;~0}  \alpha_{0~;~-\frac{3}{2}} = 1.
	\end{align}
As an example of choices satisfying the above conditions, we find
	\begin{align}
		\alpha_{ - \frac{3}{2} ~;~  0 } = \alpha_{ 0~;~ - \frac{3}{2} } = 1, ~~~~~ \chi_{ - \frac{3}{2}~;~ 0 }^* = \chi_{ 0~;~ - \frac{3}{2} } = i
	\end{align}
To see that the above data constitute a solution of the vacuum equations, we use the matrix
	\begin{align}
		M & = \left[
\begin{array}{cccccccc}
 0 & 0 & 0 & -\mu^* & 0 & 0 & 0 & \mu \\
 0 & 0 & 0 & \mu^* & 0 & 0 & 0 & \mu \\
 0 & 0 & 0 & 0 & -1 & 0 & 0 & 0 \\
 -i & 0 & 0 & 0 & 0 & 0 & 0 & 0 \\
 0 & 0 & -i & 0 & 0 & 0 & 0 & 0 \\
 0 & 0 & 0 & 0 & 0 & 0 & 1 & 0 \\
 0 & \mu & 0 & 0 & 0 & -\mu^* & 0 & 0 \\
 0 & \mu & 0 & 0 & 0 & \mu^* & 0 & 0 \\
\end{array}
\right]\in \text{SU}(8),~~~~~ \mu =  \frac{1 + i }{2},
	\end{align}	
to conjugate the entire system back to the quiver basis, i.e. the basis in which the generators $\gamma, \tau$ are manifestly direct sums of the irreducible representations described above. Mapping our solution in Jordan canonical form to the quiver basis (where for simplicity we take all $g_{(i)}$ equal as per our discussion in section \ref{sec:NAHM}), we have the following identifications:
	\begin{align}
		\vec a^t &= \vec b^t = 0, ~~~~~ \vec c = -\vec d =  \begin{bmatrix} 0 \\ -\mu^* \sqrt{\frac{3}{2}} \end{bmatrix},~~~~~ \vec e =  \vec f = \begin{bmatrix} -\mu \sqrt{\frac{3}{2}} \\ 0\end{bmatrix},~~~~~ \vec g^t =\vec h^t = \begin{bmatrix} 0&- \mu^* \sqrt{2} \end{bmatrix}\\
		\Sigma^+_{\pm{}} &= - \frac{3}{2},~~~~~ \Sigma^-_{\pm{}} = \frac{1}{2},~~~~~ \Sigma_1 = \begin{bmatrix} \frac{3}{2}  & \\ & - \frac{1}{2}  \end{bmatrix}.
	\end{align}
The above solution corresponds to the following quiver:
	\begin{align}
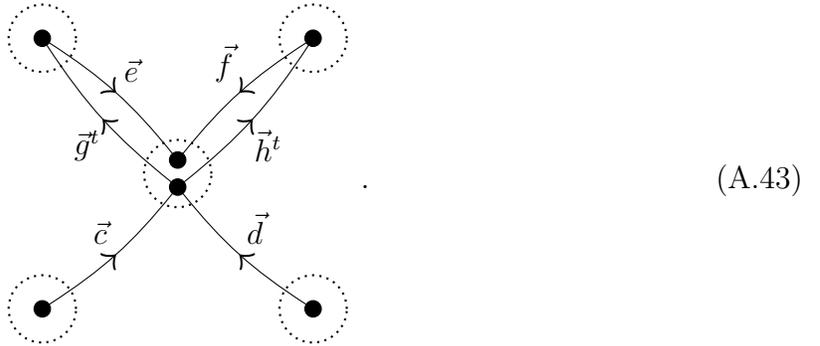
		\begin{array}{c}
			\begin{tikzpicture}[scale=1.8]
				\node[draw,circle,scale=.3,fill=black](3) at (0,0) {$1$};
				\node[draw,circle,scale=.3,fill=black](4) at (2,0) {$1$};
				\node[draw,circle,scale=.3,fill=black](1) at (0,-2) {$1$};
				\node[draw,circle,scale=.3,fill=black](2) at (2,-2) {$1$};
				\node[draw,circle,scale=.3,fill=black](6) at (1,-.9) {$1$};
				\node[draw,circle,scale=.3,fill=black](5) at (1,-1.1) {$1$};		
				\draw[->-=.5] (1) to[bend right = 9] node[above,midway]{$\vec c~~~$} (5);
				\draw[->-=.5] (2) to[bend left = 9] node[above,midway]{$~~~\vec d$} (5);
				\draw[->-=.5] (5) to[bend right = 9] node[below,midway]{$~~~\vec h^t$}(4);
				\draw[->-=.5] (4) to[bend right = 9] node[above,midway]{$\vec f~~~$} (6);
				\draw[->-=.5] (3) to[bend left = 9] node[above,midway]{$~~~\vec e$}(6);
				\draw[->-=.5] (5) to[bend left = 9] node[below,midway]{$\vec g^t~~~$}(3);
				\node[draw,circle,thick,dotted,scale=2.3] at (1,-1){};
					\node[draw,circle,thick,dotted,scale=2.3] at (2,0){};
						\node[draw,circle,thick,dotted,scale=2.3] at (0,-2){};
							\node[draw,circle,thick,dotted,scale=2.3] at (2,-2){};
								\node[draw,circle,thick,dotted,scale=2.3] at (0,0){};
			\end{tikzpicture}
		\end{array}.
	\end{align}
The solution consists of a single irreducible representation, and hence the flavor symmetry group is trivial.

\newpage

\bibliographystyle{utphys}
\bibliography{Punctures}

\end{document}